\documentclass[a4paper,11pt]{article}
\pdfoutput=1 

\usepackage{jheppub} 
\usepackage[mathscr]{euscript}
\usepackage[T1]{fontenc} 

 \usepackage{musicography}
 
\usepackage{bm}
\usepackage{verbatim}
 
 \usepackage{makecell}
 \usepackage{float}
 
 \usepackage[makeroom]{cancel}
 
 \newcommand{\bv}{ \begin{verbatim}}
 
     \newcommand{\Soft}{{ \mathsf{Soft}}}
  
    \newcommand{\bz}{{ \bar z}}
  
    \newcommand{\Split}{{ \mathsf{Split}}}
    
        \newcommand{\SV}[1]{ [ #1 ]  }
 
\newcommand{\bra}[1]{\ensuremath{\left\langle#1\right|}}

\newcommand{\be}{\begin{equation}}
\newcommand{\ee}{\end{equation}}
\newcommand{\bpm}{\begin{pmatrix}}
\newcommand{\epm}{\end{pmatrix}}

\newcommand{\PBK}[1]{\ensuremath{\begin{pmatrix}#1\end{pmatrix}}}

\newcommand{\EV}[1]{\langle #1 \rangle}
\newcommand{\beqn}{\begin{eqnarray}}
\newcommand{\eeqn}{\end{eqnarray}}

\usepackage{slashed}

\newcommand{\cJ}{\mathcal J}
\newcommand{\cO}{\mathcal O}

\newcommand{\sfp}{\mathsf p}
\newcommand{\sfk}{\mathsf k}
\newcommand{\sfh}{{\sf h}} 

\newcommand{\sfG}{\mathsf G}
\newcommand{\sfGamma}{\mathsf \Gamma}

\newcommand{\sfPhi}{{\sf\Phi}}

\newcommand{\p}{\partial}

\newcommand{\sfz}{{\sf z}}

\DeclareMathOperator{\sgn}{sgn}

\DeclareMathOperator{\Tr}{Tr}

\author[\musSixteenth{}]{Hongliang Jiang }

 \affiliation[\musSixteenth{}]{Centre for Research in String Theory, School of Physics and Astronomy,
Queen Mary University of London, Mile End Road, E1 4NS,  UK}

 \emailAdd{h.jiang@qmul.ac.uk}

\preprint{QMUL-PH-21-26} 

\title{\boldmath\Huge Celestial superamplitude in $\mathcal N=4$ SYM theory}

\abstract{Celestial amplitude is a new reformulation of momentum space  scattering amplitude  and offers a promising way for flat holography.  In this paper, we study the celestial amplitude in $\mathcal N=4$ Super-Yang-Mills (SYM) theory aiming    at understanding the role of superconformal symmetry in     celestial holography.
We first construct the superconformal generators acting on the celestial superfield which assembles all the on-shell fields in the  multiplet together in terms of  celestial variables and  Grassmann   parameters.
These generators satisfy the superconformal algebra of $\mathcal N=4$ SYM theory. We also compute the three-point and four-point celestial super-amplitude explicitly. They can be identified as the correlation functions of the celestial superfields living at the celestial sphere. We further study  the  soft and collinear limits   which give  rise to the super-Ward identity and super-OPE on  the celestial sphere, respectively. 
Our results initiate a new perspective of understanding the well-studied $\mathcal N=4$ SYM amplitude via   2D celestial conformal field theory.     
}

\begin{document} 
\maketitle
\flushbottom

 \section{Introduction}
 
 In recent years, a new formalism of scattering amplitude was developed in the study of flat holography~\cite{Pasterski:2016qvg}. In contrast to the conventional momentum basis   manifesting the translational invariance, the new formalism makes full use of the Lorentz symmetry of spacetime and expands the wave functions of particles in terms  of boost eigenstates.  Mathematically, by performing the Mellin transformation on the momentum space scattering amplitude \cite{deBoer:2003vf,Cheung:2016iub,Pasterski:2016qvg}, one obtains the so-called \emph{celestial amplitude} which behaves just like a correlation function in 2D CFT under Lorentz transformation \cite{Pasterski:2016qvg,Pasterski:2017kqt,Pasterski:2017ylz}.
 As such, this reformulation offers a promising approach to establish  flat holography    by relating the   quantum gravity scattering process in the 4D bulk Minkowski spacetime  with the observables of the 2D \emph{celestial conformal field theory}  (CCFT)  living on  the boundary celestial sphere.~\footnote{Although such a reformation  can be studied  in   arbitrary dimensions and for arbitrary fields with arbitrary masses, here we will be mainly focusing on the  massless fields in four dimensions.} This goes beyond the paradigm of AdS/CFT \cite{Maldacena:1997re} which is well-understood now after two decades of efforts. Guided by  the general holographic principle  \cite{tHooft:1993dmi,Susskind:1994vu} and considering the great success of AdS/CFT, it is  natural to ask how to concretely realize the holographic duality   for quantum gravity in the absence of cosmological constant?  Recent fruitful results     are showing  that celestial holography in terms of CCFT  opens up an interesting avenue to this question!

Besides the   connection to flat holography,  CCFT also offers a very different perspective to understand scattering amplitude itself. For example, the soft limit and collinear limits  feature  some universal properties of quantum fields and are vital for the self-consistency of S-matrix. It turns out that they just correspond to the  standard  Ward identity and OPE in CCFT \cite{Fan:2019emx,Pate:2019mfs,Nandan:2019jas,Adamo:2019ipt,Pate:2019lpp}.  CCFT   also  enables us to   succinctly characterize  the  infinite number of non-trivial symmetries  in asymptotically flat spacetime\cite{Guevara:2021abz}.
  Furthermore, CCFT   provides an advantageous language in describing the UV and IR behavior of scattering amplitude \cite{Arkani-Hamed:2020gyp}.

Despite its compelling   role,     CCFT itself remains poorly understood. On the one hand,  the celestial conformal field theory  behaves in many aspects like an ordinary two dimensional  conformal field theory. Many techniques in the ordinary CFT can be borrowed and applied to CCFT. For example, one can construct the stress tensor  \cite{Cheung:2016iub,Kapec:2016jld} and it is meaningful to discuss the operator production expansion (OPE) of various operators \cite{Fan:2019emx,Pate:2019lpp}. 
  On the other hand, CCFT   features lots of peculiarities.  In particular, the spectra of operators and the notion of  inner products, conjugation seem  to be drastically different from the conventional CFTs \cite{Pasterski:2017kqt}.  See \cite{Crawley:2021ivb,Pasterski:2021fjn} for recent discussions. 
 Nevertheless, the rich symmetries  and various self-consistency conditions already impose stringent constraints on CCFT.

 In this paper, we add one more ingredient to the CCFT by considering the quantum field theories  in 4D with superconformal symmetry.~\footnote{It is worth emphasizing  that the supersymmetry   discussed in this paper refers to the supersymmetry in the 4D bulk instead of the   conventional 2D supersymmetry. Also, the conformal symmetry here refers to the 4D bulk conformal symmetry, which is different from the   conformal symmetry on the 2D celestial sphere induced by Lorentz transformation in 4D.}   In particular, we will be interested in the   $\mathcal N=4$ super-Yang-Mills (SYM) theory.   The $\mathcal N=4$ SYM theory enjoys a huge amount of symmetries which make it tractable or  even exactly solvable in many situations. 
 Especially, the scattering amplitude in $\mathcal N=4$ SYM has been intensively studied. 
Furthermore,   $\mathcal N=4$ SYM theory provides the first and most successful example  of AdS/CFT correspondence. Given its unique role, we are naturally led to study the celestial amplitude of    $\mathcal N=4$ SYM theory and its role in flat holography.

We begin  our studies with symmetry, which is arguably one of the most important guiding principles in physics.  For celestial amplitude, a very natural question  is  how the various symmetries act on the celestial amplitude and how to  constrain the structure of  celestial amplitude by   imposing symmetries. 
 For Poincar\'e symmetry and conformal symmetry, the symmetry generators acting on   celestial amplitude are found in \cite{Stieberger:2018onx}. 
 While for supersymmetry, it was also discussed in \cite{Fotopoulos:2020bqj,Pasterski:2020pdk}. Here we generalize the discussion to superconformal symmetry and especially    the maximal    superconformal symmetry enjoyed by  $\mathcal N=4$  super-Yang-Mills  theory. 
To this aim, we make full use of the on-shell superfield formalism by introducing Grassmann variables. 
  We construct the superconformal generators in celestial superspace and checked that they indeed satisfy the superconformal algebra. These superconformal generators act on the   on-shell celestial superfield which we constructed explicitly and fulfills a representation of  the superconformal algebra. 
 These superconformal  generators can also act on the celestial super-amplitude, which is the Mellin transformation of momentum space superamplitude, and can be regarded as  the super-correlator  of on-shell celestial superfields. The superconformal symmetry thus imposes non-trivial constraints on the super-correlators: they must be invariant under the action  superconformal generators.  
   
Using Mellin transformation,  we also explicitly compute the  three-point and four-point celestial superamplitude. 
In particular, we are able to recast the celestial superamplitude  of $\mathcal N=4$   SYM theory  in a form with  not only    manifest   Lorentz symmetry  but also the obvious  dilatation symmetry and supersymmetry,   in contrast to the manifest translational invariance  and supersymmetry of momentum space superamplitude. 
By expanding the celestial superamplitude in Grassmann variables, one can obtain all the component amplitudes including the gluon amplitude.   
  
We will also study the soft limit and collinear limit of scattering amplitude. In these limits, the amplitudes diverge, featuring some universal behavior of quantum fields. For pure Yang–Mills (YM) theory, these two limits enable us to   establish the Ward identity and extract the OPE of  celestial operators, respectively \cite{Fan:2019emx,Pate:2019mfs,Nandan:2019jas,Adamo:2019ipt,Pate:2019lpp}. We generalize these discussions to our $\mathcal N=4$ SYM theory. In particular, the leading supersoft theorem in momentum space turns into  the conformally supersoft theorem on celestial sphere. Taking into account the color factors, this further leads to the Ward identity relating celestial super-correlators with and without  soft   ``super''-current  insertion. Physically, this arises from  the asymptotic large gauge transformation of gluons,     and the soft   ``super''-current is essentially the soft gluons with both helicities, generating  the ``super''-Kac-Moody   symmetry on the celestial sphere.  Again, the ``super'' here refers to the bulk supersymmetry and differs from the standard super-Kac-Moody in 2D. 
 
While in the  collinear limit, the two particles    pass through the nearby points on the celestial sphere,  thus corresponding to the coincident limit of operators. So the collinear limit enables us to derive the OPE of operators in celestial CFT \cite{Fan:2019emx,Pate:2019lpp}. Indeed, using the super-split factor for the collinear limit of SYM theory, we are able to compute the OPE between two super-operators in  $\mathcal N=4$ SYM theory.
The super-OPE encodes all the component OPEs which can be obtained easily by expanding in Grassmann variables.

We further consider  the super-OPE in the limit $\Delta\to 1, 1/2, 0, \cdots$. The resulting super-OPEs manifest themselves with  a pole in  conformal dimension $\Delta$. The operators with these special values of conformal dimension are thus  the soft  ``super''-current. In particular, when $\Delta \to 1$, the resulting OPE agrees with the one derived directly from  the soft gluon  theorem, thus providing a consistent check of CCFT.  
While for $\Delta =1/2, 0$, they correspond to soft gluinos and soft scalars, which are related to soft gluons via supersymmetry.  
 
This paper is organized as follows. In  section~\ref{scftsym}, we start with symmetry aspect  of celestial amplitude by constructing explicitly all the generators of superconformal  symmetry as well as the corresponding superfield. 
In section~\ref{CelAmp}, we compute the three and four-point celestial super-amplitude explicitly and comment on the general structure at higher points. 
In section~\ref{softlimit}, we study the soft limit of $\mathcal N=4$ SYM theory which results in the conformally soft theorem and Ward identity for celestial operators. 
In section~\ref{collimit}, we compute the OPE  of    super-operators from  the collinear limit of $\mathcal N=4$ SYM theory.  
We summarize the results of this paper in section~\ref{conclusion} and outlook possible  future directions.
We also include    appendix~\ref{app}, where we  review  the soft and collinear limit of YM and $\mathcal N=4$ SYM theory.  

\vspace{.5cm}
\emph{Noted added:} After the completion of this work, we learned that Andreas Brandhuber, Graham R. Brown, Joshua Gowdy, Bill Spence and Gabriele Travaglini, who are in the same group at QMUL, had been working independently on a very similar problem    \cite{AndiCelestial}, which overlaps with our  section~\ref{scftsym} and  \ref{CelAmp}. The results in these two papers are  consistent although using different approaches. We were not aware of the work of each other until the final stage due to Covid restrictions. 

 \section{Superconformal symmetry on the  celestial sphere} \label{scftsym}
 
 In this section, we will study the symmetry aspect  of CCFT. We will construct the superconformal generators \eqref{Rsym} -\eqref{Ssym}  in terms of the celestial sphere coordinates and Grassmann variables. They act on the celestial on-shell superfield   \eqref{celestialSuperfield}  which can be obtained as the Mellin transformation of the on-shell superfield in momentum space. These generators are shown to satisfy the superconformal algebra and impose  Ward identity constraints on the celestial superamplitude. 
 
In our discussion, we make full use of the on-shell superfield formalism which is   especially powerful for $\mathcal N=4$ SYM theory and  brings  a lot  of simplifications. Note that the role of supersymmetry was studied before in \cite{Fotopoulos:2020bqj} where the discussions were mainly based on the component form of each multiplet.

 \subsection{Kinematics on the celestial sphere }


Let us start  the celestial story with   kinematics. A massless particle travels along light-like direction in spacetime and  is described  by null momentum $p^\mu$ satisfying $p^\mu p_\mu=0$.  We can parametrize the null momentum in terms of complex coordinates $(z,\bar z)$ on the so-called celestial sphere 
 \be
p^\mu =(E,p^x, p^y, p^z)=\omega q^\mu (z,\bar z)~,
 \ee
 where $\omega>0$ and the null vector $q^\mu$ is
\be\label{qnull}
 q^\mu (z,\bar z) =\Big(1+z \bar z ,\quad  z+\bar z ,\quad -i(z-\bar z),\quad  1-z \bar z\Big)~.
\ee
The point $(z,\bar z)$ can be regarded as the point  on the celestial sphere  where the massless particle crosses,  while $\omega $ is the energy along this null direction.

Under $SL(2, \mathbb C)$, the  $(z,\bar z)$ coordinates on the celestial sphere transform  as
\be
z\to \frac{az+b}{cz+d} ~, \qquad \bar z\to \frac{\bar  a\bar  z+\bar b}{\bar c\bar z+\bar d}~, \qquad a,b,c,d \in \mathbb C~ , \quad ad-bc=1~.
\ee
As a consequence, we have
\be\label{pqTsf}
p^\mu \to \Lambda^\mu{}_\nu p^\nu~,\qquad 
q^\mu \to  |cz+d|^{-2} \Lambda^\mu{}_\nu q^\nu~,\qquad  \omega \to  |cz+d|^{ 2}  \omega~,
\ee
where $\Lambda^\mu{}_\nu $ is the associated Lorentz transformation satisfying $\Lambda^T \eta \Lambda =\eta$. This is not surprising as $SO(3,1)\simeq SL(2,\mathbb C)$. Therefore the Lorentz transformation in 4D Minkowski spacetime  induces the conformal transformation on the 2D celestial sphere, suggesting the existence of a    conformal field theory living there.

A nice property of null vector $q^\mu$ is 
\be\label{qnorm}
q^\mu(z,\bar z) q_\mu( z', \bar z') =-2 |z-  z'|^2~.
\ee

A spinning massless particle    also carries internal degrees of freedom which can be described in terms of polarization vectors. For gauge boson,   the two polarizations can be naturally chosen as 
\be
\sqrt 2 \epsilon_+^\mu (q)=\sqrt 2 \epsilon_z^\mu  =\p_z q^\mu =(\bar z,\, 1, \, -i,\, -\bar z), \qquad
\sqrt 2 \epsilon_-^\mu (q)=\sqrt 2 \epsilon_{\bar z}^\mu  =\p_{\bar z} q^\mu =(  z,\, 1, \, i,\, -  z)~,
\ee
satisfying 
\be
 \epsilon_z\cdot q= \epsilon_{\bar z}\cdot q=0,\qquad  \epsilon_z\cdot \epsilon_z= \epsilon_{\bar z}\cdot \epsilon_{\bar z}=0, \qquad 
\epsilon_z\cdot    \epsilon_{\bar z}=1~.
 \ee

We are interested in the particles involved in a scattering process where    crucially, we need to distinguish whether the particle is incoming or outgoing. So the  precise way to parametrize a null momentum  $p^\mu$ in scattering amplitude is 
\be\label{pqrelation}
 p^\mu 
 = \varepsilon \omega q^\mu (z,\bar z)=\varepsilon\omega\,\Big(1+z \bar z , \;  z+\bar z , \; -i(z-\bar z),  \; 1-z \bar z\Big)~, \qquad
 \omega>0~, \qquad z,\bar z \in \mathbb C~,
\ee
where the extra factor $\varepsilon=\sgn(p^0)$ further takes into account whether the particle is incoming ($\varepsilon=-1$) or outgoing ($\varepsilon=+1$).

In the spinor-helicity formalism, 

\be\label{pmualpha}
p^{   \alpha  \dot  \alpha} =    \sigma_\mu^{   \alpha\dot   \alpha}  p^\mu 
= \PBK{p^0+p^3  & p^1-i p^2 \\ p^1+ i p^2  &p^0 -p^3   } = \lambda^\alpha   \tilde \lambda^{\dot \alpha}  ~,
\ee
where  we choose $  \sigma_\mu^{   \alpha \dot   \alpha}  =( 1, \bm \sigma )  $.
%
%
  For real physical momentum, the two spinors are related by complex conjugation
$
 \tilde \lambda^{\dot \alpha}= \varepsilon (\lambda^\alpha)^*
$.
%
 
For the null vector introduced in \eqref{qnull}, we have 

\be
q^{   \alpha \dot \alpha}  =    \sigma_\mu^{   \alpha\dot   \alpha}  q^\mu =  \PBK{ 2 & 2\bar z \\ 2  z & 2z\bar z}
=  2 \PBK{1 \\    z}\PBK{1 &  \bar z} ~.
\ee

This suggests that we can take
\be\label{lambdaSpinor}
\lambda^\alpha\equiv \bra{p}^\alpha=\sqrt{ 2\omega } \PBK{1 \\ z} =\sqrt{ 2\omega}\bra{ q}^\alpha~,\qquad 
\tilde\lambda^{\dot \alpha} \equiv  |p]^{\dot \alpha}  =\varepsilon \sqrt{ 2\omega}   \PBK{1 \\ \bar z}=\varepsilon \sqrt{2 \omega}|q]^{\dot \alpha} ~,
\ee
where we introduce the following notation for later convenience 
\be
\bra{q}^\alpha = \PBK{1 \\ z}  ~, \qquad
|q]^{\dot \alpha} =\PBK{1\\ \bar z}~.
\ee
%
%
The angle  and square brackets of spinors are   
\beqn
\EV{ij}&=&
-2 \sqrt{\omega_i \omega_j} \; z_{ij}~, \qquad z_{ij}=z_i-z_j~,
\\ 
{}[ij] &=&
2  \varepsilon_i \varepsilon_j \sqrt{\omega_i \omega_j} \; \bar z_{ij}~, \qquad 
\bar z_{ij}=\bar z_i- \bar z_j~.
\eeqn

\subsection{Celestial amplitude}

The scattering amplitude in momentum space is given by
\be\label{momentumA}
\mathcal A_n( J_i, p^\mu_i)=A_n( J_i, p^\mu_i) \;\delta^4(\sum_i   p^\mu_i)~,\qquad p_i^\mu =\varepsilon_i \omega_i q_i^\mu~,
\ee
where $\varepsilon_i$ labels the ingoing $(-)$ and outgoing $(+)$ particle, $J_i$ is the  helicity of massless particle, and $\delta$-function is to enforce  the momentum conservation.    
 The scattering amplitude should respect the symmetry of the theory, especially the Poincar\'e symmetries which consist of  translations and Lorentz transformations.  In momentum space, this means

  \be\label{pSapceSym}
 \sum_j \varepsilon_j P_j^\mu  \mathcal A_n( J_i, p^\mu_i) =0~, \qquad  \sum_j M_j^{\mu \nu}  \mathcal A_n( J_i, p^\mu_i) =0~,
  \ee
 where the generators of translations and  Lorentz transformations act in momentum space as 
  \be
P^\mu=\varepsilon p^\mu=\omega q^\mu~, \qquad 
 M ^{\mu\nu} =p^\mu \frac{\p}{ \p p_\nu}- p^\nu \frac{\p}{ \p p_\mu}~.
  \ee
 The amplitude in the form of \eqref{momentumA}  obviously satisfies the first equation     in \eqref{pSapceSym}, while the second equation     in \eqref{pSapceSym} is not manifest    anymore. 
 So, the  momentum space scattering amplitude makes  the translation symmetry manifest but leaves the Lorentz symmetry obscure. 
 
By contrast,  the celestial amplitude makes the Lorentz symmetry  evident, while sacrificing the   manifestation of translational invariance. 
  The celestial amplitude is defined as the Mellin transformation of momentum space amplitude
 \be\label{MellinTsf}
 \mathcal M_n(\Delta_i,J_i,  z_i, \bar z_i)
 =\Big(\prod_{j=1}^n  \int_0^\infty d\omega_j\; \omega_j^{\Delta_j -1}  \Big)  \mathcal A_n( J_i, p^\mu_i)~,
 \ee
 where we parametrize the momentum of each particle as \eqref{pqrelation}.
As such,  the celestial amplitude can be regarded as a conformal correlator on the celestial sphere
\be
 \mathcal M_n(\Delta_i,J_i,  z_i, \bar z_i)=\EV{O^{\varepsilon_1}_{\Delta_1,J_1}(z_1,\bar z_1) \cdots  O^{\varepsilon_n}_{\Delta_n,J_n}(z_n,\bar z_n) }~,
\ee 
and transforms under   $SL(2,\mathbb C)$ as 
\be\label{SL2Ctsf}
 \mathcal M_n\Big(\Delta_i,J_i,  \frac{a z_i+b}{c z_i+d},  \frac{ \bar  a \bar z_i+\bar b}{\bar c \bar z_i+\bar d}\Big)
 =\Big(\prod_{j=1}^n(c z_j+d)^{\Delta_j+J_j }( \bar c \bar z_j+\bar d)^{\Delta_j-J_j } \Big) \mathcal M_n(\Delta_i,J_i,  z_i, \bar z_i)~,
\ee
where  $a,b,c,d\in \mathbb C, \; ad-bc=1$. This $SL(2,\mathbb C)$ symmetry on the celestial sphere is just the Lorentz symmetry of the spacetime. So the Lorentz symmetry is manifest, while the translation symmetry is not obvious anymore. 
To show that Poincar\'e symmetries are indeed preserved, we first need find the generators of those  symmetries in the celestial basis.  

Note that the conformal dimension and spin/helicity are related to the holomorphic and anti-holomorphic weights as
\be
\Delta=h+\bar h~, \qquad J=h-\bar h~,
\ee
where $J$  in 2d is the spin of the operator, while in 4d it is the helicity of the particle. 

In order to have a complete and  delta-function-normalizable  conformal primary wavefunctions in conformal    basis, the conformal dimension should reside in  the principal continuous series of unitary representations of $SL(2, \mathbb  C)$ \cite{Pasterski:2017kqt}
\be
\Delta=1+i\lambda ~, \qquad \lambda\in \mathbb R~.
\ee

We will also use the following   relations later:
\be
\p_\Delta  \equiv  \p_{i\lambda }\equiv -i \p_\lambda \equiv \frac12(\p_h+\p_{\bar h})~, \qquad  \p_J \equiv\frac12(\p_h-\p_{\bar h})~.
\ee

 \subsection{Poincar\'e  symmetry}

 As in the ordinary 2d CFT case, the  $SL(2,\mathbb C)$ is generated by  $L_k, \bar L_k$, $k=0,\pm 1$ whose actions on primary operator  $O_{h,\bar h}$ with conformal weights $h,\bar h$  are \cite{DiFrancesco:1997nk}
 \beqn\label{LLbaraction}
 \mathcal L_k \cdot O_{h, \bar h}(z,\bar z)&\equiv& [\mathcal L_k , O_{h, \bar h}(z,\bar z)]= L_k   O_{h, \bar h}(z,\bar z)~, \qquad
 \\
\bar {\mathcal L}_k \cdot O_{h, \bar h}(z,\bar z)&\equiv& [\mathcal {\bar L}_k , O_{h, \bar h}(z,\bar z)]=\bar L_k   O_{h, \bar h}(z,\bar z)~,
 \eeqn
 where  the differential operators $L_k, {\bar L}_k$  are given by
 \be\label{LkLkbar}
L_k= h(k+1)z^k  +z^{k+1} \p_z ~ ,\qquad \bar L_k= \bar h(k+1)\bar z^k  +z^{k+1} \p_{\bar z}  ~,
 \ee
 while $  {\mathcal L}_k, \bar {\mathcal L}_k$ should be regarded as charges in the field theory.~\footnote{More precisely, they are the mode expansion of stress tensor in the case of Virasoro symmetry.}
 They satisfy the commutation relations 
%
 \be
  [ \mathcal L_k, \mathcal L_l]= (k-l )\mathcal  L_{k+l}~, \qquad  [ \bar {\mathcal L}_k, \bar{\mathcal  L}_l]=  (k-l )\bar {\mathcal L}_{k+l}~,
 \ee 
 and
  \be\label{Lcommutator}
  [L_k, L_l]=-(k-l ) L_{k+l}~, \qquad  [ \bar L_k, \bar L_l]=-(k-l )\bar L_{k+l}~,
 \ee 
 Note that there is an extra minus sign in \eqref{Lcommutator}. The difference of the extra minus comes from the fact that  transformations compose in the opposite order when acting on the coordinates.~\footnote{More specifically, $\mathcal L_k\cdot( \mathcal L_l \cdot O) =L_l L_k O$. See the appendix of \cite{Fortin:2011nq}   for more detailed discussions on this point.} We will mostly work with $L_k, \bar L_k$ when discussing the conformal symmetry.  
The generators of  Lorentz transformations are the linear combinations of these $ SL(2,\mathbb C)$ conformal generators $L_k, \bar L_k$. 
 
 Let us now switch to translation symmetry. In momentum space, the  action of translation just multiplies the amplitude by the momentum \eqref{pSapceSym}. In  celestial basis,  the   insertion of momentum  $P^\mu_j=\omega_j q^\mu_j$ in the integrand \eqref{MellinTsf} implements a multiplication of $q^\mu_j$ and a  shift  of conformal dimension $\Delta_j\to \Delta_j+1$ (without changing the helicity). As a result, $h_j\to h_j+\frac12, \bar h_j \to \bar h_j+\frac12$. Therefore the translation generators in the celestial basis are given by \cite{Stieberger:2018onx}
 \be
 P^\mu =q^\mu e^{ \p_ \Delta}   =q^\mu e^{(\p_h +\p_{\bar h})/2}~.
 \ee
 
 Using \eqref{pmualpha}, we can rewrite it as~\footnote{Here we introduce the factor $1/2$ for simplicity, otherwise  there will be many factors of $\sqrt{2}$ flowing around. }
\be
P^{  \alpha \dot \alpha} =\frac12   \sigma_\mu^{  \alpha \dot\alpha}  P^\mu =\frac12  \sigma_\mu^{  \alpha \dot\alpha} q^\mu e^{(\p_h +\p_{\bar h})/2}
= |q]^{\dot\alpha} \bra{q}^\alpha e^{(\p_h +\p_{\bar h})/2}~,
\ee 
whose form  further suggests us to write in the product form:
 \be\label{Psymbol}
\sfp^\alpha =\bra{q }^\alpha e^{ \p_h  /2} =\PBK{1  \\ z} e^{ \p_h  /2}~, \qquad  
\tilde\sfp^{\dot\alpha}=|q]^{\dot\alpha}   e^{ \p_{\bar h}  /2}   =\PBK{1  \\  \bar z} e^{ \p_{\bar h}  /2}~ , \qquad 
P^{\dot \alpha \alpha} = \frac12   \sigma_\mu^{  \alpha\dot \alpha}  P^\mu =
  \sfp^\alpha  \tilde\sfp^{\dot\alpha}~.
   \ee
 
  \subsection{Bulk conformal  symmetry}
  
 Besides the Poincar\'e symmetry considered above, we can further consider the 4D conformal symmetry acting on bulk spacetime. Especially, the special conformal generator in the celestial basis has been worked out in \cite{Stieberger:2018onx}. As translation generators, we find special conformal generators can also be compactly rewritten as
    \be\label{Ksymbol}
\sfk^\alpha=\PBK{\p_z \\ z\p_z+2h-1 }e^{ -\p_h  /2} ~, \qquad  \tilde\sfk^{\dot\alpha}=\PBK{\p_{\bar z} \\  \bar z\p_{\bar z}+2\bar h-1 } e^{- \p_{\bar h}  /2}~ , \qquad
K^{  \alpha\dot \alpha} = \frac12   \sigma_\mu^{  \alpha\dot \alpha}  K^\mu =
  \sfk^\alpha  \tilde\sfk^{\dot\alpha}~.
   \ee
 
 Finally, the dilatation is given by
 \be\label{Dscaling}
 D= 1-(h+\bar h)  =   1-\Delta~.
 \ee

 We can then work out the full conformal algebra in this basis. Let us first show two useful relations: ($\varepsilon^{12}=-\varepsilon^{21}=1$)
  \be\label{kprelation}
 [\sfp^\alpha,\sfk^\beta]=\sfp^\alpha \sfk^\beta -  \sfk^\beta \sfp^\alpha =\varepsilon^{\alpha\beta}~, \qquad
 \sfk^\alpha \sfp^\beta-\sfk^\beta \sfp^\alpha=-2(h-1)\varepsilon^{\alpha\beta}~,
 \ee
 and its conjugate
   \be
 [\tilde\sfp^{\dot \alpha}, \tilde \sfk^{\dot \beta}]=\tilde\sfp^{\dot \alpha}  \tilde \sfk^{\dot \beta} -  \tilde \sfk^{\dot \beta} \tilde\sfp^{\dot \alpha} =\varepsilon^{\dot\alpha \dot\beta}~, \qquad
\tilde\sfk^{\dot \alpha} \tilde\sfp^{\dot \beta}- \tilde\sfk^{\dot \beta} \tilde\sfp^{\dot \alpha}=-2( \bar h-1)\varepsilon^{\dot\alpha \dot\beta}~.
 \ee
 
 We can then construct  
  \be
 L^{\alpha\beta}=   \frac12(\sfk^\alpha \sfp^\beta+  \sfk^\beta \sfp^\alpha)~, \qquad
  L^{\dot\alpha\dot\beta}=   \frac12(\tilde  \sfk^{\dot\alpha}\tilde \sfp^{\dot \beta}+  \tilde\sfk^{\dot\beta} \tilde\sfp^{\dot\alpha})~,
 \ee
which enables us to write
 \be\label{Lkprelation}
 \sfk^\alpha \sfp^\beta=L^{\alpha \beta}-(h-1)\varepsilon^{\alpha\beta}~, \qquad
\sfp^\alpha \sfk^\beta=L^{\alpha \beta} +h\varepsilon^{\alpha\beta}~,
 \ee
 and likewise for $  L^{\dot\alpha\dot\beta}$.
 
 It turns out that these are just the $L_k,\bar L_k$ in \eqref{LkLkbar}. More precisely, we have 
 \be
 L_{11}=L_{-1}~, \quad L_{12}=L_{21}=L_0~, \quad L_{22}=L_1~, \quad
  L_{\dot 1\dot 1}=\bar L_{-1}~, \quad L_{\dot 1\dot 2}=L_{ \dot 2\dot 1}=\bar L_0~, \quad L_{\dot 2\dot 2}=\bar L_1~.
 \ee
 So $ L_{\alpha\beta},     L_{\dot\alpha\dot\beta} $ are just the   generators of $SL(2, \mathbb C)$.~\footnote{The Lorentz generators of  4D spacetime can be written as $M^{\mu\nu} =(\sigma^{\mu\nu})_{\alpha\beta} L^{\alpha\beta} +(\bar \sigma^{\mu\nu})_{\dot\alpha\dot\beta}L^{\dot\alpha\dot\beta}$  where $\sigma^{\mu\nu} \propto \sigma^{[\mu}\sigma^{\nu]} $, $\bar\sigma^{\mu\nu} \propto   \bar \sigma^{[\mu}  \bar\sigma^{\nu]}$.}

 
 With these generators, one can show that they indeed satisfy the conformal algebra $\mathfrak{so}(4,2)$:
  \beqn
  [K^{\alpha\dot\alpha},P^{\beta\dot\beta}] 
 &=&
 \varepsilon^{\alpha\beta} L^{\dot\alpha\dot\beta}
+ \varepsilon^{\dot\alpha\dot\beta}L^{\alpha\beta} 
+ D\varepsilon^{\alpha\beta} \varepsilon^{\dot\alpha\dot\beta} ~,
\\ 
{}  [D,K^{\alpha\dot\alpha}]&=&-K^{\alpha\dot\alpha}~,
\\  
{} [D,P^{\alpha\dot\alpha}]&=&P^{\alpha\dot\alpha}~,
\\ 
{} [L^{\alpha\beta},L^{\gamma\delta}] &=&\varepsilon^{\alpha\delta} L^{\gamma\beta} -  \varepsilon^{\gamma\beta} L^{\alpha\delta}~, \qquad
[L^{\dot\alpha\dot\beta},L^{\dot\gamma\dot\delta }] =\varepsilon^{\dot\alpha\dot\delta} L^{\dot\gamma\dot\beta} -  \varepsilon^{\dot\gamma\dot\beta} L^{\dot\alpha\dot\delta}~,
\\ 
{} [L^{\alpha\beta},P^{\gamma\dot\gamma}]  &=&-\varepsilon^{\gamma\beta} P^{\alpha\dot\gamma}+\frac12\varepsilon^{\alpha\beta}P^{\gamma\dot\gamma} ~, \qquad
[L^{\dot\alpha\dot\beta},P^{\gamma\dot\gamma}]  =\varepsilon^{\dot\alpha\dot\gamma} P^{   \gamma \dot\beta}-\frac12\varepsilon^{\dot\alpha\dot\beta}P^{\gamma\dot\gamma} ~,
\\ 
{} [L^{\alpha\beta},K^{\gamma\dot\gamma}]  &=&-\varepsilon^{\gamma\beta} K^{\alpha\dot\gamma}+\frac12\varepsilon^{\alpha\beta}K^{\gamma\dot\gamma} ~, \qquad
[L^{\dot\alpha\dot\beta},K^{\gamma\dot\gamma}]  =\varepsilon^{\dot\alpha\dot\gamma} K^{   \gamma \dot\beta}-\frac12\varepsilon^{\dot\alpha\dot\beta}K^{\gamma\dot\gamma}  ~.
\eeqn

 
For any field theory  with conformal symmetry, the amplitude should respect these symmetries. So we have 
  \beqn \label{AmpWI1}
  && \sum_j  \varepsilon_j P_j^{\alpha\dot\beta}  \mathcal M_n( h_,\bar h_i,  z_i, \bar z_i  )=  \sum_j    L_j^{\alpha\beta}  \mathcal M_n( h_,\bar h_i,  z_i, \bar z_i  )= \sum_j     L_j^{\dot\alpha\dot\beta}   \mathcal  M_n( h_,\bar h_i,  z_i, \bar z_i  ) 
\nonumber\\ =&& \label{AmpWI2}
   \sum_j  \varepsilon_j K_j^{\alpha\dot\beta}  \mathcal M_n( h_,\bar h_i,  z_i, \bar z_i  ) = \sum_j  D_j  \mathcal M_n( h_,\bar h_i,  z_i, \bar z_i  )=0 ~,
    \eeqn
    where the subscript $j$ means the action of the generators on the $j$-th particle. 

 For comparison, we also write down the conformal generators in the spinor-helicity basis \cite{Henn:2014yza}: 
\beqn\label{spinGenerator}
&&\mathsf P^{\alpha \dot\alpha}=\lambda^\alpha \tilde \lambda^{\dot \alpha}~, \qquad 
\mathsf K^{\alpha \dot\alpha} = \frac{\p}{\p \lambda_{ \alpha} } \frac{\p}{\p \lambda_{\dot\alpha} }   ~,\qquad
\mathsf D=\frac12 \lambda^\alpha  \frac{\p}{\p\lambda^\alpha } + \frac12 \tilde \lambda^{\dot\alpha } \frac{\p}{\p \tilde \lambda^{\dot\alpha }} +1~,
\\&&\label{spinGenerator2}
\mathsf L^{\alpha\beta}=\frac12\Big( \lambda^\alpha \frac{\p}{\p\lambda_ \beta }  +\lambda^\beta \frac{\p}{\p\lambda_ \alpha }   \Big)~ ,
\qquad
\mathsf L^{\dot \alpha \dot \beta}=\frac12\Big( \tilde \lambda^{\dot \alpha} \frac{\p}{\p\tilde\lambda_{\dot \beta} }  
+\tilde \lambda^{\dot \beta} \frac{\p}{\p\tilde\lambda_{\dot \alpha} }   \Big) ~.
\eeqn

   \subsection{Bulk superconformal  symmetry and $\mathcal N=4$ SYM }
   Now we turn to supersymmetry. Our main interest in this paper is $\mathcal N=4$ SYM theory which enjoys the maximally superconformal symmetry $\mathfrak{psu}(2, 2|4)$ in 4D. So to study the celestial amplitude in   $\mathcal N=4$ SYM theory, we also would like to find the generators of the superconformal symmetry.  We will first review the super-amplitude in  spinor-helicity basis and then generalize to the  celestial basis. 
   
   Let us first recall the field content of  $\mathcal N=4$ SYM theory which includes the spin-1 gluon, four spin-1/2 gluinos and six real scalars.  All of them transform in the adjoint representation of the gauge group. For scattering amplitude, we only need to consider the on-shell degrees of freedom (d.o.f).  There are $8_B+8_F$ on-shell degrees of freedom and we list them in the following table~\ref{n4dof}:
 \begin{table}[H]
\begin{center}
\begin{tabular} {|c|c|c|c|c|c|}   \hline
particle & $\mathsf G^+$ & $\mathsf G^-$ &  $\mathsf \Phi_{AB}=-\mathsf \Phi_{BA}$ & $\mathsf \Gamma_A$ & $\bar{\mathsf\Gamma}^A $\\ \hline
 helicity  $J$& $+1$&$-1$&0 & $+\frac12 $&$-\frac12$ \\ \hline
 on-shell d.o.f. &1  &1 &6 & 4&4 \\ \hline
 $SU(4)_R$&  singlet ($\bf 1$) & singlet ($\bf 1$)  & anti-symmetric ($\bf 6$) &   fundamental ($\bf 4$) & anti-fundamental ($\bar{\bf 4}$)   \\ \hline
\end{tabular} 
\end{center}
\label{n4dof}
\caption{On-shell degrees of freedom of $\mathcal N=4$ SYM. Note that the color index for gauge group is  suppressed.}
\end{table}

\subsubsection{Superconformal symmetry in spinor-helicity basis}

It is convenient to   introduce the anti-commuting Grassmann variables $\eta^A$ with $A=1, \cdots,4$. Then
the on-shell degrees of freedom  of  $\mathcal N=4$ SYM theory can be packaged into an on-shell  superfield  as follows \cite{Henn:2014yza}:
\beqn\label{Phisuperfield}
\mathsf\Psi(p, \eta)&=&\mathsf G^+(p)+\eta^A \mathsf\Gamma_A(p)
+\frac{1}{2!}\eta^A\eta^B \mathsf\Phi_{AB}(p) \nonumber
\\&&
+\frac{1}{3!}\epsilon_{ABCD} \eta^A\eta^B \eta^C \bar{\mathsf\Gamma}^{D}(p)
+\frac{1}{4!}\epsilon_{ABCD} \eta^A\eta^B \eta^C\eta^D \mathsf G^- (p)~.
\eeqn
Note each component field  in this superfield   has different helicities.  We can formally assign $\eta^A$ helicity $\frac12$, then each term in the on-shell superfield \eqref{Phisuperfield}  carries the same helicity. Mathematically, we can introduce the following generators
\be\label{Jgenerator}
\mathsf \Omega=-\frac12 \lambda^\alpha  \frac{\p}{\p\lambda^\alpha }+\frac12 \tilde \lambda^{\dot\alpha } \frac{\p}{\p \tilde \lambda^{\dot\alpha }} +\frac12 \eta^A \p_A ~, \qquad \p_A=\frac{\p}{\p\eta^A} ~,
\ee
whose action on the superfield is then $\mathsf\Omega \mathsf  \Psi=\mathsf\Psi$.

The Poincar\'e  and conformal supersymmetry generators acting on the  superfield  \eqref{Phisuperfield}  are given by \cite{Henn:2014yza}: 
 \beqn \label{Qsusy}
&&
\mathsf Q^{\alpha\, A}= \lambda^\alpha\eta^A ~, \qquad\tilde {\mathsf Q}^{\dot \alpha }_A=\tilde \lambda^{\dot \alpha}\frac{\p}{\p\eta^A} ~,
\\&&  \label{Ssusy}
\mathsf S^{\alpha}_{ A}=\lambda^\alpha  \frac{\p}{\p\eta^A}~ , \qquad\tilde  {\mathsf S}^{\dot \alpha  \,A } =\tilde \lambda^{\dot \alpha} \eta^A~.
\eeqn
Besides, the $R$-symmetry generators are given by \cite{Henn:2014yza}:
 \be\label{Rsymm}
\mathsf R^A{}_B=\eta^A \p_B -\frac14 \delta^A_B \,\eta^C\p_C ~.
\ee
Together with those in \eqref{spinGenerator} and \eqref{spinGenerator2}, they generate the full  superconformal symmetry of $\mathcal N=4$ SYM.

The superamplitude provides a compact way to write down the amplitude of SYM theory:
\be\label{superamp}
\mathscr A_n(p_i,\eta_i) =\EV{\mathsf\Psi(p_1,\eta_1)\cdots \mathsf\Psi(p_n,\eta_n) }~.
\ee
Performing the $\eta$ expansion and comparing both sides, one can get the  scattering  amplitude of all the component fields.  The superamplitude is severely constrained by the superconformal symmetry of SYM theory. Especially this implies $\sum_j O_j \mathscr A_n(\lambda_i,\tilde\lambda_i,\eta_i) =0$ where $O\in\{ \mathsf L,\mathsf P,\mathsf D,\mathsf K,\mathsf Q,\mathsf S,\mathsf R\} $. We will not spell out all the details here which can be found in \cite{Henn:2014yza}. 

 We can then read  off the $Q$-supersymmetry variation of the component  field defined by 
\footnote{ One can also consider the $\bar Q$-supersymmetry variation:
$
 \delta_{\bar \epsilon } \Phi(p, \eta) =\bar \epsilon_{\dot\alpha}^{ A} Q^{\dot\alpha}_{ A} \Phi( p,\eta) \nonumber
$
as well as the $S, \tilde S$-supersymmetry variation in a similar way.
}
\be
\delta_\epsilon  \mathsf\Psi(p, \eta) =\epsilon_{\alpha A} Q^{\alpha A} \mathsf\Psi( p,\eta)~.
\ee
Expanding both sides in components, we have
\beqn
\delta_\epsilon  \mathsf\Psi(p, \eta) &=& \delta_\epsilon  \mathsf G^+(p)+\eta^A \delta_\epsilon   \mathsf\Gamma_A(p)
+\frac{1}{2!}\eta^A\eta^B\delta_\epsilon   \mathsf\Phi_{AB}(p) +\cdots~,
\\
 \epsilon_{\alpha A} Q^{\alpha A} \mathsf\Psi( p,\eta)&=&
 \epsilon_{\alpha A} \lambda^\alpha \Big(
  \eta^A \mathsf G^+(p)+\eta^A\eta^B \mathsf\Gamma_B(p)
+\frac{1}{2!}\eta^A\eta^B\eta^C \mathsf\Phi_{BC}(p)+\cdots 
\Big)~,
\eeqn
which gives the supersymmetry variations of various component fields: 
%
\beqn\label{susyvariation}
  \delta_\epsilon  \mathsf G^+ =0~, \quad  
   \delta_\epsilon   \mathsf\Gamma_A
  =  \epsilon_{\alpha A} \lambda^\alpha   \mathsf G^+=-\EV{\epsilon_A \lambda} \mathsf G^+~,
\quad
\delta_\epsilon \mathsf\Phi_{AB}=  -\EV{\epsilon_A \lambda}   \mathsf\Gamma_B+ 
 \EV{\epsilon_B \lambda}   \mathsf\Gamma_A~, \quad \cdots ~.
\eeqn

 \subsubsection{Quasi-supersymmetry generators in celestial basis}
 Now we would like to perform a similar construction in the celestial basis. First we need to find the supersymmetry generators in terms of the celestial coordinates. Motivated by the construction  in  the spinor-helicity basis  in \eqref{Qsusy} and \eqref{Ssusy}, it is natural   to guess that  the   Poincar\'e  and conformal supersymmetry generators take the following form: 
 \beqn\label{QsusyCB}
&& Q^{\alpha\, A}= \sfp^\alpha\eta^A ~, \qquad\tilde Q^{\dot \alpha }_A= \tilde \sfp^{\dot \alpha}\frac{\p}{\p\eta^A}~,
\\
&& S^{\alpha}_{ A}=\sfk^\alpha  \frac{\p}{\p\eta^A} ~, \qquad\tilde S^{\dot \alpha  \,A } =\tilde \sfk^{\dot \alpha} \eta^A~,
\label{KsusyCB}
\eeqn
%
%
 %
where $ \sfp, \tilde \sfp,\sfk, \tilde \sfk $ are given in \eqref{Psymbol} and \eqref{Ksymbol}.
In particular, it is easy to check that they   satisfy the supersymmetry algebra as desired: 
\be
\{ Q^{\alpha \, A} ,\tilde Q^{\dot \alpha}_{ B} \}=\delta^A_B P^{\alpha\dot\alpha}~,\qquad
\{ S^{\alpha}_{ A},\tilde S^{\dot \alpha\, B} \}=\delta^B_A K^{\alpha\dot\alpha}~.
\ee
%

Besides, we also have the R-symmetry generators which take  the same form as \eqref{Rsymm}: 
\be
R^A{}_B=\eta^A \p_B -\frac14 \delta^A_B \,\eta^C\p_C~, 
\ee
as well as
\be
\Omega=h-\bar h +\frac12 \eta^A\p_A~.
\ee
 
 We can further compute 
\beqn
\{Q^{\alpha\, A},S^{\beta}_{ B}  \}&=&   \varepsilon^{\alpha\beta} R^A{}_B+\delta^A_B L^{\alpha\beta} 
-\frac12\varepsilon^{\alpha\beta} \delta^A_B (D- \Omega+1)~, 
\\
\{\bar Q^{\dot\alpha}_B, \bar S^{\dot\beta\,A}  \}  &=& 
-   \varepsilon^{\dot\alpha\dot\beta} R^A{}_B+\delta^A_B L^{\dot\alpha\dot\beta} 
-\frac12\varepsilon^{\dot\alpha\dot\beta} \delta^A_B (D+ \Omega-1)~,
\eeqn 
where we used the identities \eqref{kprelation} and \eqref{Lkprelation}. Note that when acting on on-shell superfield or super-amplitude, we have $\Omega=1$. 

One can also check all the rest of the commutators. Indeed they generate the $\mathfrak {psu}(2,2|4)$ superconformal algebra, which is the symmetry of $\mathcal N=$ SYM theory. However, it turns out that the generators constructed in this way are a bit subtle as they don't  act on the physical on-shell superfield as we expected. For this reason,   the generators constructed above  will be called  {quasi}-superconformal generators. We will explain this point in  detail in the following.

\subsubsection{Celestial on-shell superfield and celestial superamplitude }
Now we also would like to have the analog of the superamplitude \eqref{superamp} in the celestial basis. For this purpose, we need to  generalize the notion of superfield in \eqref{Phisuperfield}. 
The superconformal generators can  then  act on the celestial on-shell superfield and realize a representation of the superconformal algebra. 
It turns out that this brings a subtlety which will be resolved at the end of this section. 

 \subsubsection*{Celestial on-shell  celestial superfield from Mellin transformation}
A natural way to obtain the superfield in celestial basis  is to  perform the Mellin transformation of   \eqref{Phisuperfield}:
 \be
  \widehat \Psi_\Delta(z,\bar z,  \eta)=\int_0^\infty d\omega\; \omega^{\Delta-1} \mathsf\Psi(\varepsilon\omega q^\mu(z,\bar z), \eta)~.
 \ee
 Since the $\eta$-expansion and Mellin integral commute, we can perform Mellin transformation component by component. 
 Each component has the same conformal dimension, but different helicity. The results can thus be compactly written as
\beqn\label{celestialSuperfield}
\widehat\Psi_{h,\bar h}(z,\bar z,   \eta)&=&\mathsf G^+(z,\bar z, h,\bar h)+\eta^A \mathsf\Gamma_A(z,\bar z, h-\frac14,\bar h+\frac14)
+\frac{1}{2!}\eta^A\eta^B \mathsf\Phi_{AB}(z,\bar z, h-\frac12,\bar h+\frac12)
\nonumber\\&&
+\frac{1}{3!}\epsilon_{ABCD} \eta^A\eta^B \eta^C \bar{\mathsf\Gamma}^{D}(z,\bar z, h-\frac34,\bar h+\frac34)
+\frac{1}{4!}\epsilon_{ABCD} \eta^A\eta^B \eta^C\eta^D \mathsf G^- (z,\bar z, h-1,\bar h+1) 
\nonumber\\
\eeqn
with the extra condition $h-\bar h=1$. Then obviously every component indeed has the same conformal dimension $\Delta=h+\bar h$ and its own correct helicity (note that the Grassmann variable $\eta$ has $\Delta=0$ and $J=1/2$).
 
 The superamplitude  is given by 
 \be\label{celAMP}
\widehat{\mathscr M}_n(h_i,\bar h_i, z_i, \bar z_i, \eta_i) =\EV{ \widehat\Psi_{h_1,\bar h_1}(z_1,\bar z_1,   \eta_1)   \cdots  \widehat\Psi_{h_n,\bar h_n}(z_n,\bar z_n,   \eta_n) }~.
\ee

 However, this on-shell superfield is not appropriate for the supercharge $Q,\bar Q$ constructed in  \eqref{Qsusy}. From the supersymmetry variation in  \eqref{susyvariation}, we obtain
  \be\label{deltaGamma}
   \delta_\epsilon   \mathsf\Gamma_A(\Delta)=\int d \omega \;\omega^{\Delta-1}    \delta_\epsilon   \mathsf\Gamma_A(\omega)
 = \int d \omega\; \omega^{\Delta-1}  (z-\zeta_A)  \sqrt \omega\mathsf G^+ (\omega)
 =  (z-\zeta_A) \mathsf G^+ (\Delta+\frac12)~,
 \ee
 where we used $\epsilon^\alpha_{  A}=\frac{1}{\sqrt{2}} (1,\zeta_A)^T$.~\footnote{Note that $\epsilon_{\alpha A} \in \mathbb {CP}^1\simeq S^2$. So $\zeta_A$ transforms under the $SL(2,\mathbb C)$ transformation in the same way as $z$. } As a result, we have  
 \be\label{Qsusytsf}
    \delta_\epsilon   \mathsf\Gamma_A(h,\bar h )= (z-\zeta_A) \mathsf G^+  (h+\frac12, \bar h ),\qquad \bar h=h-\frac12~.
 \ee
 However, applying the supercharge \eqref{QsusyCB} to the \eqref{celestialSuperfield},  the supersymmetry variation rule  $\delta_\epsilon \widehat \Phi= \epsilon_{\alpha A}  Q^{\alpha A} \widehat \Phi $  leads to $\delta_\epsilon   \mathsf\Gamma_A(h,\bar h)= (z-\zeta_A) \mathsf G^+  (h+\frac34, \bar h-\frac14 )$ which is not consistent with the above result \eqref{Qsusytsf}.   To reproduce  the expected  supersymmtry variations, one needs to modify the rule as follows:
  \be
 \delta_\epsilon \widehat \Psi =\epsilon_{\alpha \, A}  \widehat Q^{\alpha\, A} \widehat \Psi~, \qquad
  \delta_{\bar \epsilon }\widehat  \Psi= \bar\epsilon_{\dot\alpha}^A  \widehat {\bar Q}^{\dot\alpha}_ A \widehat \Psi~.
 \ee
 where 
  \be\label{newQsusy}
 \widehat Q^{\alpha \, A} =e^{\frac{\p_{\bar h} - \p_{  h}}{4} }Q^{\alpha \, A} =\PBK{1\\z} e^{\frac{\p_{\bar h} + \p_{  h}}{4} }  \eta^A 
  ~,\qquad
  \widehat {\bar Q}^{\dot\alpha }_{A} =e^{\frac{\p_{  h} - \p_{ \bar  h}}{4} }Q^{\alpha }_{A} =\PBK{1\\ \bar z} e^{\frac{\p_{\bar h} + \p_{  h}}{4} } \p_A~.
 \ee
 These supercharges are just those used in  \cite{Fotopoulos:2020bqj,Pasterski:2020pdk}.   However, at this moment,  it is not obvious how to similarly modify the rest of generators in order to fulfill a representation acting   on \eqref{celestialSuperfield}. This will be achieved at the end of this subsection. Before that, let us stick to the quasi-superconformal generators that we constructed in the previous subsection and try to find  a way to fulfill its representation. 
 
\subsubsection*{Quasi-on-shell superfield }
It turns out that we can introduce the following   superfield 
 \beqn\label{Phisuperfield2}
 \Psi_{h,\bar h}(z,\bar z,    \eta)&=&\mathsf G^+(z,\bar z, h,\bar h)+\eta^A \mathsf\Gamma_A(z,\bar z, h,\bar h)
+\frac{1}{2!}\eta^A\eta^B \mathsf\Phi_{AB}(z,\bar z, h,\bar h) \nonumber
\\&&
+\frac{1}{3!}\epsilon_{ABCD} \eta^A\eta^B \eta^C \bar{\mathsf\Gamma}^{D}(z,\bar z, h,\bar h)
+\frac{1}{4!}\epsilon_{ABCD} \eta^A\eta^B \eta^C\eta^D \mathsf G^- (z,\bar z, h,\bar h)~.
\eeqn
Applying the  Poincar\'e  supersymmetry generators \eqref{QsusyCB} to the above superfielld,  the  corresponding supersymmetry transformation rules
  \be
 \delta_\epsilon   \Psi=\epsilon_{\alpha \, A}    Q^{\alpha\, A}   \Psi ~, \qquad 
   \delta_{\bar \epsilon }   \Psi= \bar\epsilon_{\dot\alpha}^A    Q^{\dot\alpha}_ A   \Psi 
 \ee
indeed give  the right variation \eqref{Qsusytsf}.

However, there is a subtle point related to the superfield \eqref{Phisuperfield2}. Different components there can not become on-shell simultaneously due to  their different helicities but the same value of $h-\bar h$.  For this reason, we will call \eqref{Phisuperfield2} \emph{quasi-on-shell  superfield}. 
Nevertheless, this  subtlety does not bring any physical pathologies because we  always need to pick exactly one component in the multiplet  for each external leg to get the     amplitude in a physical scattering process.
So we can  regard \eqref{Phisuperfield2} as a formal sum where the conformal weights  $h,\bar h$ are arbitrary a priori. The on-shell condition for $h-\bar h$  is imposed appropriately only after the component field is specified. 
 
 The corresponding superamplitude  is given then by 
 \be
 {\mathscr M}_n(p_i,\eta_i) =\EV{   \Psi_{  h_1,\bar h_1}(z_1,\bar z_1,   \eta_1)   \cdots    \Psi_{h_n,\bar h_n} (z_n,\bar z_n,   \eta_n) }~.
\ee

 To obtain the component field amplitude, we first  perform the $\eta$-expansion, choose a specific component for each external leg, and finally impose the helicity constraint for each $h_i-\bar h_i$. 
 
 \subsubsection*{Celestial on-shell superfield,  celestial superamplitude and  superconformal generators }
 
 In the above discussions, we have constructed the superconformal generators \eqref{QsusyCB}, \eqref{KsusyCB} whose action on the quasi-on-shell superfield  \eqref{Phisuperfield2} fulfills the representation of the superconformal algebra. Although we have argued that the  quasi-on-shell superfield  \eqref{Phisuperfield2}  does not bring physical pathologies, its interpretation is still obscure conceptually. 
 In the following, we will construct a new set of superconformal generators. They act on the celestial on-shell-superfield \eqref{celestialSuperfield} and also realize    a representation of the superconformal algebra.

 By comparing the component expansion of two  superfields \eqref{celestialSuperfield} and  \eqref{Phisuperfield2}, it   is not difficult to find that   they  can be  related as follows: 
  \be\label{2PhiRelation}
\widehat{  \Psi} = e^{-\frac12 \eta^A\p_A \p_J}  \Psi   
= e^{-\frac14 \eta^A\p_A (\p_h-\p_{\bar h})}  \Psi ~ , \qquad   \Psi = e^{\frac12 \eta^A\p_A \p_J}\widehat \Psi   ~,
 \ee
where $ e^{-\frac12 \eta^A\p_A \p_J}$ implements the shift of  $h$ and $\bar h$ according to the number of $\eta$'s. 
 
We have shown that $ \Psi$ fulfills the representation of the superconformal algebra with generators 
$\{P,L,K,D,Q,S,R \}$. To find the generators acting on the on-shell superfield $\widehat\Psi$,  we can naturally modify generators in the following way
 \be
\widehat O=  e^{-\frac12 \eta^A\p_A \p_J} Oe^{\frac12 \eta^A\p_A \p_J}= e^{-\frac14 \eta^A\p_A(\p_h-\p_{\bar h})} O  e^{ \frac14 \eta^A\p_A(\p_h-\p_{\bar h})}~,
 \ee
 where $O\in \{P,L,K,D,Q,S,R \}$  is any superconformal generators we constructed before. Now $\widehat O $ acts on the on-shell superfield  $\widehat \Psi$ in \eqref{celestialSuperfield}.  By construction, this should also fulfill the      superconformal representation and  in particular the generators should satisfy the superconformal algebra. In the followings, we will   write down the generators $\{ \widehat P,\widehat L,\widehat K,\widehat D,\widehat Q,\widehat S,\widehat R \}$ explicitly and verify these statements. 
 
Let us start by observing the following useful relation: 
  \be\label{usefulrelation}
 e^{- \eta^A\p_A \p_x} \eta^{A_1} \cdots \eta^{A_k} \p_{B_1} \cdots \p_{B_l}f(x)  e^{  \eta^A\p_A \p_x}
 =f(x-   \eta^A\p_A) \eta^{A_1} \cdots \eta^{A_k}  \p_{B_1} \cdots \p_{B_l}e^{ (l-k)\p_x}~,
 \ee
 which can be proved by inserting $ 1=e^{  \eta^A\p_A \p_x}  e^{- \eta^A\p_A \p_x} $ between $\p_{B_l} $ and $f(x)$.~\footnote{Note the relation
$  
 [\eta^A\p_A,\eta^B]=\eta^B, \qquad [\eta^A\p_A,\p_B ]=-\p_B,\qquad
  [\eta^A\p_A,\eta^B \p_B]=0
$.}
%
%
This immediately implies that $e^{-  \eta^A\p_A \p_x} \eta^B\p_C  e^{   \eta^A\p_A \p_x}= \eta^B\p_C$. So we have%
  \be \label{Rsym}
 \widehat R^A{}_B=R^A{}_B= \eta^A \p_B -\frac14 \delta^A_B \eta^C \p_C~,
 \ee
 and 
  \beqn
\widehat \Omega&=& 
 e^{-\frac14 \eta^A\p_A(\p_h-\p_{\bar h})} (h-\bar h+\frac12 \eta^A\p_A)  e^{ \frac14 \eta^A\p_A(\p_h-\p_{\bar h})}
=
(h -\frac14 \eta^A\p_A )- (\bar h+\frac14 \eta^A\p_A)+\frac12 \eta^A\p_A
\nonumber \\& =&
h-\bar h=J~.
 \eeqn
 So we identically have $\widehat\Omega\widehat \Phi=\widehat\Phi$ as $h-\bar h=1$ for  $\sfG^+$ gluon. 
 
 It is also easy to find that 
  \be\label{Doperator}
\widehat D=D=1-h-\bar h~.
 \ee
 
The special conformal generators are now:   
 \be
\widehat K^{ \dot \alpha \alpha}  
 =e^{-\frac14 \eta^A\p_A(\p_h-\p_{\bar h})} 
\tilde\sfk^{\dot\alpha} \sfk^\alpha  e^{ \frac14 \eta^A\p_A(\p_h-\p_{\bar h})}
=\hat{\tilde\sfk}^{ \dot\alpha} \, \hat \sfk^{\alpha }~,
 \ee
 where
 \beqn
\hat\sfk^{ \alpha} =e^{-\frac14 \eta^A\p_A \p_h } \sfk^\alpha  e^{ \frac14 \eta^A\p_A \p_h }
&=&  \PBK{\p_z \\ z\p_z+2h-\frac12 \eta^A\p_A -1 }e^{ -\p_h  /2} ~,
\\
\hat{\tilde\sfk}^{ \dot\alpha} =e^{ \frac14 \eta^A\p_A \p_{\bar h} } \sfk^\alpha  e^{- \frac14 \eta^A\p_A \p_{\bar h} }
&=&  \PBK{\p_{\bar z} \\ \bar z\p_{\bar z}+2\bar h+\frac12 \eta^A\p_A -1 }e^{ -\p_{\bar h}  /2} ~.
 \eeqn
 It is worth comparing the special conformal generators here with that in \eqref{Ksymbol}: the  only difference is the shift of $h,\bar h$ by $\mp\frac14 \eta^A\p_A$.  Looking at the superfield \eqref{celestialSuperfield}, the shift is indeed reasonable because it ensures that the special conformal generators act  on each component in a correct way with its right value of $h,\bar h$.
  
 The Lorentz generators become: 
 \beqn\label{newLoretz}
\widehat  L_{\alpha\beta} &=& 
  \frac12 ( \hat \sfk^\alpha  \sfp^\beta+ \hat \sfk^\beta \sfp^\alpha)~, \qquad
\widehat   L_{\dot\alpha\dot\beta} =
   \frac12 (\hat {\tilde  \sfk}^{\dot\alpha}\tilde \sfp^{\dot \beta}+\hat{  \tilde\sfk}^{\dot\beta} \tilde\sfp^{\dot\alpha})~.
 \eeqn

 The translation generators remain the same: 
 \be
 \widehat P^{ \dot \alpha \alpha} =   P^{ \dot \alpha \alpha} = {\tilde\sfp}^{ \dot\alpha} \,   \sfp^{\alpha }~.
 \ee
 The $Q$-supersymmetry generators are 
 \beqn\label{Qcharge}
 \widehat Q^{\alpha\, A}&=&e^{-\frac14 \eta^A\p_A(\p_h-\p_{\bar h})} \sfp^\alpha \eta^A  e^{ \frac14 \eta^A\p_A(\p_h-\p_{\bar h})}
= \sfp^\alpha \eta^Ae^{-\frac14 (\p_h-\p_{\bar h})}=\bra{q}^\alpha e^{\frac12 \p_\Delta}\eta^A~,
\\
 \widehat{\bar Q}^{\dot\alpha} _A &=&e^{-\frac14 \eta^A\p_A(\p_h-\p_{\bar h})} \tilde\sfp^{\dot\alpha} \p_ A  e^{ \frac14 \eta^A\p_A(\p_h-\p_{\bar h})}
= {\tilde\sfp}^{\dot\alpha} \p_Ae^{ \frac14 (\p_h-\p_{\bar h})}=|q] ^{\dot\alpha} e^{\frac12 \p_\Delta}\p_A~ ,
 \eeqn
 where we used the relation \eqref{usefulrelation}. The $Q$-supercharges constructed now  agree with  \eqref{newQsusy}. 
 
 Making full use of  \eqref{usefulrelation}, one also finds the $S$-supersymmetry generators:
  \beqn
\widehat S^\alpha_A&=&\hat\sfk^{\alpha} \p_A e^{\frac14(\p_h-\p_{\bar h})}
=  \PBK{\p_z \\ z\p_z+2h-\frac12 \eta^B\p_B -1 }  \p_Ae^{-\frac14(\p_h   + \p_{\bar h})}~,
\\ \label{Ssym}
\widehat S^{\dot\alpha \, A}&=&\hat{\tilde\sfk}^{\dot\alpha} \eta^A e^{\frac14(-\p_h+\p_{\bar h})}
=  \PBK{\p_{\bar z} \\ \bar z\p_{\bar z}+2\bar h+\frac12 \eta^B\p_B -1 }  \eta^Ae^{-\frac14(\p_h   + \p_{\bar h})}~.
 \eeqn
 
These are all the superconformal generators acting on the on-shell superfield $\widehat \Phi$. As we said, by construction they satisfy the superconformal algebra.  Now we  can compute some  commutators of these generators and verify this claim.
 
 First of all, it is easy to see that 
 \be
 \{  \widehat Q^{\alpha\, A},  \widehat{\bar Q}^{\dot\alpha} _B \}= \delta^A_B\widehat P^{\alpha\dot\alpha}~,\qquad
   \{\widehat S ^\alpha_A ,\widehat  S ^{\dot\alpha \, B}\} 
  =\hat{\tilde\sfk}^{ \dot\alpha} \sfk^{ \alpha} \{  \p_A e^{\frac14(\p_h-\p_{\bar h})}, \eta^B e^{\frac14(-\p_h+\p_{\bar h})} \}= \delta^B_A\widehat K^{\alpha\dot\alpha}~,
 \ee
 where we used the identity 
  \be
 [AB,CD]_\pm=[A,C]BD+CA[B,D]_\pm \qquad \text{if} \qquad [B,C]=[A,D]=0~,
 \ee
and noticed  that      
  \be
 [\hat\sfk^{ \alpha} ,  \eta^Ae^{-\frac14(\p_h -\p_{\bar h})   } ]=0,\qquad
  [\hat{\tilde\sfk}^{ \dot\alpha} ,  \p_Ae^{-\frac14  (\p_{\bar h}  -\p_h)} ]=0,\qquad
   [\hat \sfk^{ \alpha} , \hat{\tilde\sfk}^{\dot\alpha} ]=0~.
 \ee

 The identities \eqref{kprelation} and \eqref{Lkprelation} are generalized to:
 \beqn
 [\sfp^\alpha, \hat \sfk^\beta] &=&   \varepsilon^{\alpha\beta}~, \qquad\qquad
\hat  \sfk^\alpha \sfp^\beta- \hat \sfk^\beta \sfp^\alpha=-2(h-\frac14 \eta^A\p_A-1)\varepsilon^{\alpha\beta}~,
\\
\hat \sfk ^\alpha \sfp^\beta&=&\widehat L ^{\alpha \beta}-(h-\frac14 \eta^A\p_A-1)\varepsilon^{\alpha\beta}~, \qquad
  \sfp^\alpha\hat \sfk ^\beta=\widehat L^{\alpha \beta} +(h-\frac14 \eta^A\p_A)\varepsilon^{\alpha\beta}~.
 \eeqn
 They enable us to check 
 \beqn
  \{  \widehat Q^{\alpha\, A},    \widehat S ^\beta_B \}
  &=&[\sfp^\alpha,\hat \sfk^\beta] \eta^A\p_B+\hat \sfk^\beta\sfp^\alpha \delta^A_B
   = \varepsilon^{\alpha\beta} \eta^A\p_B+\Big( \widehat L ^{\alpha \beta}+(h-\frac14 \eta^A\p_A-1)\varepsilon^{\alpha\beta}\Big) \delta^A_B
\nonumber\\&=&
 \widehat L ^{\alpha \beta} \delta^A_B+ \varepsilon^{\alpha\beta} \widehat R^A_B -
 \frac12 \varepsilon^{\alpha\beta}(\widehat D-\widehat \Omega+1)~.
 \eeqn
 All the rest of the commutators can also be checked similarly and indeed they generate the superconformal algebra $\mathfrak{psu}(2,2|4)$.

 The superamplitude should be invariant under the action of these symmetries, so it must satisfy the Ward identities for all the  superconformal generators:
 %
%
    \be\label{AmpWI3}
\sum_j    \Big\{ \varepsilon_j \widehat P_j,  \;  \widehat  L_j^{\alpha\beta}  , \; \widehat  L_j^{\dot\alpha\dot\beta},\;  \varepsilon_j \widehat K_j^{\alpha\dot\beta},\;  \widehat   D_j ,\;  ( \widehat  Q  ^{\alpha  }_A)_j    , \; \varepsilon_j ( \widehat {\bar Q}^{\dot\alpha\, A} )_j, \;  ( \widehat  R^A_B)_j   ,\; ( \widehat  S  ^{\alpha  }_A)_j   , \; \varepsilon_j  (\widehat {\bar S}^{\dot\alpha\, A} )_j  
      \Big\}\widehat {\mathscr M}_n( h_i,\bar h_i,  z_i, \bar z_i ,\eta_i )=0~,
    \ee
    in addition to the helicity constraint $\widehat \Omega_j\widehat {\mathscr M}_n( h_i,\bar h_i,  z_i, \bar z_i ,\eta_i )=\widehat{\mathscr M}_n( h_i,\bar h_i,  z_i, \bar z_i ,\eta_i )  $ for each $j$. 
    %
  
Especially, the Lorentz invariance is manifest. By generalizing \eqref{SL2Ctsf}, the celestial  superamplitude transforms under $SL(2,\mathbb C)$    as
    \beqn\label{SL2Ctsf2}
\widehat{\mathscr M}_n(h_i,\bar h_i, z_i,\bar z_i,\eta_i)
& =&
\Big(\prod_{j=1}^n(c z_j+d)^{-2 h_j +\frac12 \eta_j^A \p_{j A} }( \bar c\bar  z_j+\bar  d)^{-2\bar  h_j -\frac12 \eta_j^A \p_{j A} }   \Big)  \widehat{  \mathscr  M}_n\Big(h_i,\bar h_i,  \frac{a z_i+b}{c z_i+d},  \frac{ \bar  a \bar z_i+\bar b}{\bar c \bar z_i+\bar d},\eta_i\Big)
\nonumber \\   
  & =&
 \Big(\prod_{j=1}^n(c z_j+d)^{-2 h_j   }( \bar c\bar  z_j+\bar  d)^{-2\bar  h_j   }   \Big)
\widehat{\mathscr M}_n\Bigg(h_i,\bar h_i,  \frac{a z_i+b}{c z_i+d},  \frac{ \bar  a \bar z_i+\bar b}{\bar c \bar z_i+\bar d},
\Big( \frac{ c z_i+d   }{ \bar c\bar z_i+\bar d } \Big)^{\frac12} \eta_i  \Bigg)~.
 \qquad   \quad  
\eeqn
The second equality follows because the celestial superamplitude is the sum of different component terms where each term contains specific number of $\eta$'s; shifting the power in $ c z_j+d , \bar c \bar z_j+\bar d $ according to the number of $\eta_j$ just implements the transformation of $\eta_j$ itself under  $SL(2,\mathbb C)$, which is also consistent with its conformal weight  $(h_\eta,\bar h_\eta)=(\frac14,-\frac14)$. One can also check explicitly that all the superconformal generators leave the conformal weights invariant as desired, otherwise different  terms  in the sum in \eqref{AmpWI3} carry different weights due to the action of generators on different particles    and thus break  the    $SL(2,\mathbb C)$   covariance.

 In the following, we will be mainly using the on-shell superfield $\widehat\Psi$ as its physical meaning is clearer and has manifest  $SL(2,\mathbb C)$   covariance.  To ease the notation,  we will ignore the hat on the superfield  \eqref{celestialSuperfield}, the superamplitude \eqref{celAMP}   and generators \eqref{Rsym} -\eqref{Ssym}.
 

 \section{Celestial superamplitude in $\mathcal N=4$ SYM theory }\label{CelAmp}
 
In this section, we will compute the (color-ordered) celestial superamplitude in $\mathcal N=4$ SYM theory, focusing especially on the three-point and four-point case. This is done by performing a Mellin transformation on  the  momentum space superamplitude.  As we will see explicitly, the resulting celestial superamplitude indeed can be regarded as the correlators of celestial superoperators and it has also manifest conformal invariance and supersymmetry. 
 The computation technique here closely follows that in   pure YM case \cite{Pasterski:2017ylz}. We will first discuss the four-point amplitude, and then the three-point case. 
 
  \subsection{Celestial amplitude in YM theory}

 Before considering the superamplitude in SYM, let us first review the YM case. Due to the conformal symmetry, the YM amplitude has the  following scaling property  
 \be\label{YMscaling}
\mathcal A_n (L \omega_i, z_i,\bar z_i) =L^{-n} \mathcal A_n( \omega_i, z_i,\bar z_i)~, \qquad
 \mathcal A_n( \omega_i, z_i,\bar z_i) =A_n(p_i)\delta^4(  P)  ~.
\ee

Then the celestial amplitude is given by the Mellin transform \eqref{MellinTsf}. For four-particle scattering $1,2\to 3,4$, actually the scaling property of YM amplitude ensures that the celestial amplitude takes the following form \cite{Pasterski:2017ylz}:
\beqn
\mathcal M_4&=&\prod_{i=1}^4 \int_0 ^\infty d\omega_i \;\omega_i^{\Delta_i-1 }A_4(\omega_i, z_i, \bar z_i) \delta(\sum_i \varepsilon_i \omega_i q_i)
 \\&=& 
\frac{ \pi}{2}\delta(\sum_i \lambda_i)    \Theta(\chi-1)\delta( \chi-\bar\chi )
\Big( \prod_{i=1}^4  s_{*i}^{i\lambda_i}  \Big) \; \frac{1}{|z_{13} z_{24}|^2 }
  A_4(  s_{*j},z_j,\bar z_j)  ~,
  \label{4particelCamp}
\eeqn
where $\Theta$ equals to 0 (1) for negative (positive) argument and
%
\be
s_{*1}= s_{*4}\frac{z_{24}\bar z_{34}}{z_{12}\bar z_{13}}, \qquad
 s_{*2}= -s_{*4} \frac{z_{34}\bar z_{14}}{z_{23}\bar z_{12}},\qquad
 s_{*3}= -  s_{*4}  \frac{z_{24}\bar z_{14}}{z_{23}\bar z_{13}},\qquad
\ee
%
%
%
%
 %
The cross-ratio is defined as 
\be
 \chi=\frac{z_{12}z_{34}}{z_{13}z_{24}}~, \qquad  \bar  \chi=\frac{ \bar z_{12} \bar z_{34}}{\bar z_{13}\bar z_{24}}~.
\ee
The different factors in \eqref{4particelCamp} can be explained as follows.  The factor $\delta(\sum_i \lambda_i)$ accounts for the scaling property \eqref{YMscaling} as well as the following integral
\be
\int_0^\infty d\omega \, \omega^{i\lambda-1}=2\pi \delta(\lambda)~.
\ee
Physically this is just the manifestation of the conformal symmetry: it ensures that the Ward identity \eqref{AmpWI3} for dilatation \eqref{Dscaling} is identically satisfied.  The factor $\delta( \chi-\bar\chi )$ just comes from momentum conservation which ensures that in the four particle scattering process the momenta of for particles should lie on the same plane. The intersection of such a plane with celestial sphere forces the cross-ratio to be real. 

For concreteness, let us    consider the four-gluon  scattering with $1^-, 2^-\to 3^+,4^+$, namely $\varepsilon_{1,2}=-1,\varepsilon_{3,4}=1$ and $J_{1,2}=-1,J_{3,4}=+1$. The   color-ordered amplitude in momentum space  is  given by 
\be 
\mathcal A_4(\omega_j,z_i,\bar z_i)=A_4 (p_i) \delta^4(\sum_i p_i)=\frac{\EV{12}^3}{\EV{23}\EV{34}\EV{41}}  \delta^4(\sum_i \varepsilon_i \omega_i q_i)=\frac{\omega_1 \omega_2}{\omega_3\omega_4} \frac{z_{12}^3}{ z_{23}z_{34}z_{41}}\delta^4(\sum_i \varepsilon_i \omega_i q_i)~.
\ee
Plugging into \eqref{4particelCamp}, we get the corresponding celestial amplitude \footnote{Actually, there is  also an overall  factor $(-1)^{i\lambda_2+i\lambda_3}$  which will be suppressed here and below.}  
\be \label{22gluonAmp}
\mathcal M_4=-  \frac{ \pi}{2 }  \Theta(\chi-1)\delta( \chi-\bar\chi )   \delta\Big( \sum_j \lambda_j  \Big)
K\Big( \frac{i\lambda_1}{2},\frac{i\lambda_2}{2},1+\frac{i\lambda_3}{2},1+\frac{i\lambda_4}{2} ;1+
 \frac{ i\lambda_1}{2} ,1+ \frac{i\lambda_2}{2} ,  \frac{i\lambda_3}{2} , \frac{i\lambda_4 }{2} \Big) 
 \frac{ \chi ^{\frac53}}{(1-\chi)^\frac13}
\ee
where we introduce the kinematic factor $K$
\be\label{Kfactor}
K(h_1, h_2, h_3, h_4; \bar h_1,\bar  h_2, \bar h_3, \bar h_4)=    \prod_{i<j} z_{ij}^{\frac{h}{3}-h_i -h_j}\bar z _{ij}^{\frac{\bar h}{3}-\bar h_i -\bar h_j}, \qquad h=\sum_i h_i, \qquad \bar h=\sum_i h_i~.
\ee
So the four-gluon celestial amplitude indeed takes the form of four-point function in CFT with weights $h_i,\bar h_i$ indicated above. 

\subsection{Four-point celestial superamplitude in SYM}
Now we move to the amplitude in SYM theory. The color-ordered superamplitude is given by \cite{Henn:2014yza}
\be\label{4ptMHVsuper}
\mathscr A_4(\omega_j,z_j,\bar z_j)=\frac{ \delta^4(P) \delta^8(\mathsf Q)}{\EV{12} \EV{23}\EV{34}\EV{41}} =\frac{ \delta^4(P) \delta^8(\mathsf Q)}{(-2)^4 \omega_1 \omega_2\omega_3\omega_4} \frac{1}{ z_{12}  z_{23}z_{34}z_{41}}
\ee
where
\be
\delta^8(\mathsf Q)=\delta^8( \sum_{i=1}^4 \lambda_i^\alpha\eta^A_i)=\prod_{A=1}^4 \sum_{\substack{ i,j=1\\i<j \\}}^4 \EV{ij} \eta_i^A\eta_j^A=\prod_{A=1}^4 \sum_{\substack{ i,j=1\\i<j \\}}^4 -2 \sqrt{\omega_i\omega_j}z_{ij}  \eta_i^A\eta_j^A
\ee
The delta function $\delta^4(P) ,\delta^8(\mathsf Q)$ guarantees that the amplitude conserves the momentum       and is supersymmetric invariant.   
The celestial amplitude can be again obtained through Mellin transformation.  In general for $\mathscr A_4(\omega_i,z_i,\bar z_i) =\mathsf   A_4(\omega_i,z_i,\bar z_i)\delta^4(P)\delta^8(\mathsf Q)
$, we have
\beqn
  {\mathscr M}_4 &=&\prod_{k=1}^4 \int_0 ^\infty d\omega_i \;\omega_k^{\Delta_k-1 } \mathscr A_4(\omega_i,z_i,\bar z_i)
\\
&=&\prod_{k=1}^4 \int_0 ^\infty d\omega_k \;\omega_k^{\Delta_k-1 } {\mathsf A}_4(\omega_i,z_i,\bar z_i)  \delta(\sum_i \varepsilon_i \omega_i q_i) \prod_{A=1}^4  \sum_{\substack{ i,j=1\\i<j \\}}^4 -2 \sqrt{\omega_i\omega_j}z_{ij}  \eta_i^A\eta_j^A
\\
&=& \prod_{A=1}^4 \sum_{\substack{ i,j=1\\i<j \\}}^4 -2 z_{ij}  \eta_i^A\eta_j^A  e^{\frac{\p_{\Delta_i}+\p_{\Delta_j}}{2}}\prod_{k=1}^4 \int_0 ^\infty d\omega_k \;\omega_k^{\Delta_k-1 } {\mathsf A}_4(\omega_i,z_i,\bar z_i)  \delta(\sum_i \varepsilon_i \omega_i q_i) 
\\
&=& \prod_{A=1}^4  \sum_{\substack{ i,j=1\\i<j \\}}^4 z_{ij}  \eta_i^A\eta_j^A  e^{\frac{\p_{\Delta_i}+\p_{\Delta_j}}{2}}   \Bigg[  \frac{ \pi}{2}\delta(\sum_k    \lambda_k+4i)    \Theta(\chi-1)\delta( \chi-\bar\chi )  \Big( \prod_{k=1}^4  s_{*k}^{i\lambda_k}  \Big) \; \frac{(-2)^4 \mathsf  A_4(  s_{*j},z_j,\bar z_j)  }{|z_{13} z_{24}|^2 }  \Bigg]~.\qquad\qquad
 \eeqn
For the superamplitude in \eqref{4ptMHVsuper}, we can easily compute 
\beqn
&&\Big( \prod_{k=1}^4  s_{*k}^{i\lambda_k}  \Big) \; \frac{ \mathsf A_4(  s_{*j},z_j,\bar z_j)   }{|z_{13} z_{24}|^2 }
 \delta(\sum_k \lambda_k+4i)    
 \nonumber \\&=& 
-  K( {\sf h}_i; \bar  {\sf h}_i)  \;
  \chi ^{\frac13} (1-\chi) ^{\frac13}\;  \delta(\sum_k \lambda_k+4i)  ~, \qquad\qquad
  \sfh_i \equiv  1+\frac{i\lambda_i}{2}~,\qquad \bar \sfh_i \equiv  \frac{i\lambda_i}{2}~,
\eeqn
where we use the fact that the cross-ratio is real  $\chi =\bar \chi$.  

Therefore the four-point celestial super-amplitude \eqref{celAMP} can be written as:
 \beqn
   {\mathscr M}_4  &=& \EV{  \Psi_{h_1,\bar h_1}(z_1,\bar z_1,   \eta_1)   \cdots   \Psi_{h_4,\bar h_4}(z_4,\bar z_4,   \eta_4) }
 \\
&=&- \frac{ \pi}{2}\delta(\sum_k    \lambda_k )    \Theta(\chi-1)\delta( \chi-\bar\chi )   \prod_{A=1}^4  \sum_{\substack{ i,j=1\\i<j \\}}^4  z_{ij}  \eta_i^A\eta_j^A  e^{\frac{\p_{\Delta_i}+\p_{\Delta_j}}{2}}   \Bigg[    \chi ^{\frac13} (1-\chi) ^{\frac13}\; 
K( {\sf h}_i; \bar  {\sf h}_i)  \Bigg] ~. \qquad \quad
\label{4ptCamp2}
\eeqn
The celestial super-amplitude can be regarded as the  conformal  correlation function of superoperators on the celestial sphere.~\footnote{Note that the fermionic delta function  \eqref{deltaQ} has vanishing weights under $  SL(2,\mathbb C)$. Indeed, obviously $e^{\p_{\Delta_i}/2}$ has weight $(h,\bar h) =(\frac14,  \frac14)$, $\eta_i$ has weight $(h,\bar h) =(\frac14, - \frac14)$, and $\bra{q_i}^\alpha$ has weight $(h,\bar h) =(-\frac12,0)$ as can be seen from \eqref{pqTsf}.
 }
  The shift operation $e^{\p_{\Delta_i}/2}$ in the fermionic delta function seems to be unconventional; but actually it can be replaced with the multiplication factor $K^{-1}e^{\p_{\Delta_i}/2} K=\prod_{k\neq i} (z_{ik} \bar z_{ik})^{-\frac14} $ once acting on   the square bracket.

Performing the $\eta$ expansion, we get all the component amplitude.  For example, consider the $\prod_{A=1}^4 \eta_1^A\eta_2^A$ term in the expansion. Using   
\beqn
&& 
 \Big(z_{12}e^{\frac{\p_{i \lambda_1}+\p_{i \lambda_2}}{2}} \Big)^4 
 K\Big(1+\frac{i\lambda_i}{2}; \frac{i\lambda_i}{2}\Big)\;
   \chi ^{\frac13} (1-\chi) ^{\frac13}\;
  \\&=&
K\Big( \frac{i\lambda_1}{2},\frac{i\lambda_2}{2},1+\frac{i\lambda_3}{2},1+\frac{i\lambda_4}{2} ;
 \frac{1+i\lambda_1}{2} ,1+ \frac{i\lambda_2}{2} ,  \frac{i\lambda_3}{2} , \frac{i\lambda_4 }{2} \Big) 
 \frac{ \chi ^{\frac53}}{(1-\chi)^\frac13}  ~,
\eeqn
one obtains 
 \beqn
 && \EV{  \Phi_{h_1,\bar h_1}(z_1,\bar z_1,   \eta_1)   \cdots   \Phi_{h_n,\bar h_n}(z_n,\bar z_n,   \eta_n) } \Big|  _{ \eta_1^4\eta_2^4 }
\nonumber \\
&=& - \frac{ \pi}{2}\delta(\sum_k    \lambda_k )    \Theta(\chi-1)\delta( \chi-\bar\chi )   K\Big( \frac{i\lambda_1}{2},\frac{i\lambda_2}{2},1+\frac{i\lambda_3}{2},1+\frac{i\lambda_4}{2} ;
 \frac{1+i\lambda_1}{2} ,1+ \frac{i\lambda_2}{2} ,  \frac{i\lambda_3}{2} , \frac{i\lambda_4 }{2} \Big) 
 \frac{ \chi ^{\frac53}}{(1-\chi)^\frac13}  ~.
 \qquad\qquad
\eeqn
This agrees with the   four-gluon celestial superamplitude in \eqref{22gluonAmp}.   
 
 Similarly, we can consider the four gluino amplitude by picking up the term $\eta_1^1\eta_3^1 \prod_{A=2}^4 \eta_2^A \eta_4^A$:
  \beqn
 && \EV{  \Phi_{h_1,\bar h_1}(z_1,\bar z_1,   \eta_1)   \cdots   \Phi_{h_n,\bar h_n}(z_n,\bar z_n,   \eta_n) } \Big|  _{ \eta^1_1\eta_3^1 \prod_{A=2}^4 \eta_2^A \eta_4^A }
\nonumber \\
&=& - \frac{ \pi}{2}\delta(\sum_k    \lambda_k )    \Theta(\chi-1)\delta( \chi-\bar\chi )   K\Big(  h_i, \bar h_i \Big) 
 \frac{1}{\chi(1-\chi)^\frac13}  ~,
 \qquad\qquad
\eeqn
where the conformal weights are 
\beqn
&& h_1=\frac34+\frac{i\lambda_1}{2}~, \qquad  \bar h_1=\frac14+\frac{i\lambda_1}{2}~, \\
&& h_2=\frac14+\frac{i\lambda_2}{2}~, \qquad  \bar h_2=\frac34+\frac{i\lambda_2}{2}~, \\
&& h_3=\frac34+\frac{i\lambda_3}{2}~, \qquad  \bar h_3=\frac14+\frac{i\lambda_3}{2}~, \\
&& h_4=\frac14+\frac{i\lambda_4}{2}~, \qquad  \bar h_4=\frac34+\frac{i\lambda_4}{2}~.
\eeqn
This corresponds to the four-gluino  scattering $1 ^{+\frac12}, 2 ^{-\frac12} \to 3 ^{+\frac12} ,4 ^{-\frac12} $. The result can be checked to agree with the one obtained from  Mellin transformation directly. All the rest of component amplitudes can be obtained in a similar way.
 
 \subsection{Structure at higher points}
 Just like the super-amplitude  \eqref{4ptMHVsuper} written in a form with manifest translation symmetry and supersymmetry, the celestial superamplitude \eqref{4ptCamp2} can also be written as 
 \be\label{M4compact}
 \mathscr M_4=\delta(D) \delta^8(Q)   \Bigg[ -\frac{\pi}{2} \Theta(\chi-1) \delta(\chi-\bar \chi)   \chi ^{\frac13} (1-\chi) ^{\frac13}\; 
K( {\sf h}_i; \bar  {\sf h}_i)  \Bigg]~,
 \ee
where the dilatation and supercharges are
  \beqn \label{deltaDQ}
\delta(D) &=& \delta(\sum_i  D_i) =\delta(\sum_i \lambda_i) ~,\\
\label{deltaQ}
\delta^8(  Q)&=&\delta^8\Big( \sum_{ i}  \bra{q_i}^\alpha\eta^A_i e^{\frac12 \p_{\Delta_i}}\Big)=\prod_{A=1}^4 \sum_{\substack{  \\i<j \\}}  z_{ij} \eta_i^A\eta_j^Ae^{\frac{\p_{\Delta_i} +\p_{\Delta_j}}{2} }~.
\eeqn
They follow from our superconformal generators defined in       \eqref{Doperator} and  \eqref{Qcharge}.
The celestial amplitude written in this form has thus manifest dilatation symmetry and supersymmetry. 
This structure can be generalized to higher points.

The $n$-point    superamplitude in momentum space  takes   the following form \cite{Henn:2014yza}
 \be\label{nptMHVsuper}
\mathscr A_n =\frac{ \delta^4(P) \delta^8(\mathsf Q)}{\EV{12} \EV{23}\cdots \EV{n1}} H_n(\lambda_i, \tilde\lambda_i, \eta_i), \qquad n\ge 4~,
\ee
where
\be\label{Hexpansion}
H_n(\lambda_i, \tilde\lambda_i, \eta_i)=H_n^{(0)} +H_n^{(4)}+H_n^{(8)}+\cdots+ H_n^{(4n-16)}, \qquad  H_n^{(l)}\sim \mathcal O(\eta^l)~.
\ee
Each term in the expansion corresponds to MHV, NMHV, $ \cdots$, $\overline {\text{MHV}}$, respectively. For MHV, $H_n^{(0)} =1$.

Similarly, the corresponding celestial superamplitude is supposed to have the following general structure: 
\be\label{nptMHVCAmp}
        {\mathscr M}_n =\delta(D )  \delta^8(  Q) \; \mathcal F_n(h_i,\bar h_i,z_i,\bar z_i,\eta_i) ~,
    \ee
where $\delta(D),\delta^8(Q)$ are given in \eqref{deltaDQ} and \eqref{deltaQ}. 
The function $\mathcal F_n$ can be regarded as an $n$-point function in CCFT and  transforms covariantly under $SL(2,\mathbb C)$.~\footnote{More precisely, both $ {\mathscr M}_n $ and $\mathcal F_n$ obey  \eqref{SL2Ctsf2}. Note $\delta(D),\delta^8(Q)$  are left invariant under $SL(2,\mathbb C)$. } It also  admits an expansion as \eqref{Hexpansion} and each term can be written as the product of the kinematic factor (namely the  $n$-point generalization of \eqref{Kfactor}), the function of cross-ratios, and Grassmann variables. The explicit form of $\mathcal F_n$ can be obtained from Mellin transformation as in the YM case.~\footnote{Considering the rich symmetry of $\mathcal N=4$ SYM, it would be interesting to see whether it is possible to bootstrap   $\mathcal F_n$ without performing Mellin transformation.} The expression \eqref{nptMHVCAmp} reduces to \eqref{4ptCamp2} when $n=4$ as expected. So the general structure in \eqref{nptMHVCAmp} has not only the manifest Lorentz invariance, but also  the manifest dilatation symmetry and supersymmetry.

In principle, we can compute all the higher-point celestial superamplitudes explicitly. For pure YM theory, the higher point was considered in \cite{Schreiber:2017jsr}.   Instead  of doing tedious higher point computation, we will consider the three-point celestial amplitude in the rest of this section.

\subsection{Three-point celestial superamplitude in SYM}
The three-point   amplitude vanishes  on-shell identically in Minkowski signature due to the kinematic  constraints. But we can define the three-point celestial  in the (2,2) split signature. As such, the $z_{ij}$ and $\bar z_{ij}$ become  real and independent variables. And there are two helicity configurations for three-point amplitude, MHV and $\overline{\mathrm {MHV}}$.

 For three-point gluon amplitude  $\mathcal A_3(\omega_j,z_j,\bar z_j)= A_3(\omega_j,z_j,\bar z_j)  \delta(\sum_i \varepsilon_i \omega_i q_i)$,
one can obtain the celestial amplitude by performing   the Mellin transformation and using agin its scaling property. The momentum conservation leads to two branches.  In particular, on the branch where $z_{ij}\neq 0$, the celestial amplitude reads
\footnote{Here we used the equation:
\be\nonumber
\delta(\sum_i \varepsilon_i s_i q_i) \delta(\sum_i s_i-1) \Big|_{z_{ij}\neq 0}
=\frac{\delta(\bar z_{13}) \delta(\bar z_{23})}{4 s_1 s_2 s_3 D_3^2} \prod_{i=1}^3\delta(s_i-s_{*i})
=\frac{\delta(\bar z_{13}) \delta(\bar z_{23})  \sgn(z_{23}z_{31}) }{4  z_{23}z_{31} s_3  } \prod_{i=1}^3\delta(s_i-s_{*i})~.
\ee
}
\be
\mathcal M_3=
\frac{\pi}{2} \delta(\sum \lambda_i) \prod_{i=1}^3  s_{*i}^{i\lambda_i} 
A_3( s_{*j},z_j,\bar z_j)\frac{\delta(\bar z_{13}) \delta(\bar z_{23})  \sgn(z_{23}z_{31}) }{   z_{23}z_{31}  \; s_{*3} }  \Theta\Big( \frac{ \varepsilon_1}{ \varepsilon_3} \frac{z_{23} }{  z_{12}} \Big)  \Theta\Big( \frac{ \varepsilon_2}{ \varepsilon_3} \frac{z_{31} }{  z_{12}} \Big) ~,
\ee
where 
\be\label{s123}
s_{*1}= \varepsilon_1 \frac{z_{23}}{D_3}, \qquad
s_{*2}= \varepsilon_2 \frac{z_{31}}{D_3}, \qquad
s_{*3}= \varepsilon_3 \frac{z_{12}}{D_3}, \qquad
D_3=  \varepsilon_1 z_{23}+ \varepsilon_2 z_{31}+ \varepsilon_3 z_{12}~.
\ee
The celestial amplitude   on  the other    branch   $\bar z_{ij}\neq 0$ is similarly obtained by exchanging $z_{ij}\leftrightarrow \bar z_{ij}$. 

For MHV amplitude  $1^- 2^-   3^+$ in YM theory, we have
\be
\mathcal A_{--+} (\omega_j,z_j,\bar z_j) = \frac{\EV{12}^3}{\EV{23}\EV{31}}\delta^4(P)=-2 \frac{\omega_1\omega_2}{\omega_3} \frac{z_{12}^3}{z_{23}z_{31}}\delta^4(\sum_i  \varepsilon_i \omega_i q_i^\mu) ~.
\ee
The corresponding celestial amplitude is thus given by 
\be\label{3ptMHVcelestialAmp}
\mathcal M_{--+}  =-\pi \delta(\sum_i\lambda_i) \frac{z_{12}^3}{z_{23}^2z_{13}^2} \Big( \frac{ \varepsilon_1}{ \varepsilon_3} \frac{z_{23}}{z_{12}}\Big)^{1+i\lambda_1}\Big( \frac{ \varepsilon_2}{ \varepsilon_3} \frac{z_{31}}{z_{12}}\Big)^{1+i\lambda_2}
\sgn(z_{23}z_{31}) \delta(\bar z_{13})\delta(\bar z_{23})   
\Theta\Big( \frac{ \varepsilon_1}{ \varepsilon_3} \frac{z_{23} }{  z_{12}} \Big)  \Theta\Big( \frac{ \varepsilon_2}{ \varepsilon_3} \frac{z_{31} }{  z_{12}} \Big) ~.
\ee

The three-point MHV superamplitude in SYM theory is 
 
\be 
\mathscr A^{\text{MHV}}_3(\omega_j,z_j,\bar z_j)=\frac{ \delta^4(P) \delta^8(\mathsf Q)}{\EV{12} \EV{23}\EV{3 1}} =\frac{ \delta^4(P) \delta^8(\mathsf Q)}{ (-2)^3\omega_1 \omega_2\omega_3 } \frac{1}{ z_{12}  z_{23}z_{31} }~,
\ee
where
\be
\delta^8(\mathsf Q)=\delta^8( \sum_{i=1}^3 \lambda_i^\alpha\eta^A_i)
=\EV{12}^4 \delta^4 \Big( \eta_1 - \frac{\EV{23}}{\EV{12}}\eta_3  \Big) 
 \delta^4 \Big( \eta_2 - \frac{\EV{31}}{\EV{12}}\eta_3  \Big) 
 =\prod_{A=1}^4 \sum_{\substack{ i,j=1\\i<j \\}}^4 \EV{ij} \eta_i^A\eta_j^A~,
\ee
where we present two equivalent representations. 

Written in terms of the celestial variables and replacing    $ \omega_i \to e^{\p_{\Delta_i}}$ in the Mellin integral, $\delta^8(\mathsf Q)$ becomes 
\be
(-2)^4 z_{12}^4 \; \delta^4 \Big( \eta_1 e^{ \frac12 \p_{\Delta_1}} - \frac{z_{23}}{z_{12}}\eta_3 e^{ \frac12 \p_{\Delta_3}}  \Big) 
 \delta^4 \Big( \eta_2 e^{ \frac12 \p_{\Delta_2}}- \frac{z_{31}}{z_{12}}\eta_3  e^{ \frac12 \p_{\Delta_3}}\Big) 
  =\prod_{A=1}^4 \sum_{\substack{ i,j=1\\i<j \\}}^4 -2 z_{ij}e^{\frac{1}{2}(\p_{\Delta_i}+\p_{\Delta_j})} \eta_i^A\eta_j^A~.
\ee

Then one can derive the  three-point celestial superamplitude which generally takes the form
\beqn
\mathscr M_3 &=&
2\pi\delta(\sum_i \lambda_i) \frac{\delta(\bar z_{13}) \delta(\bar z_{23})  \sgn(z_{23}z_{31}) }{4  z_{23}z_{31}  \; s_{*3} } \Theta\Big( \frac{ \varepsilon_1}{ \varepsilon_3} \frac{z_{23} }{  z_{12}} \Big)  \Theta\Big( \frac{ \varepsilon_2}{ \varepsilon_3} \frac{z_{31} }{  z_{12}} \Big) z_{12}^4  (-2)^4\;\mathsf A_3( s_{*j},z_j,\bar z_j)
\nonumber\\& &
\times  \delta^4 \Big( \eta_1 e^{ \frac12 \p_{\Delta_1}} - \frac{z_{23}}{z_{12}}\eta_3 e^{ \frac12 \p_{\Delta_3}}  \Big) 
 \delta^4 \Big( \eta_2 e^{ \frac12 \p_{\Delta_2}}- \frac{z_{31}}{z_{12}}\eta_3  e^{ \frac12 \p_{\Delta_3}}\Big) 
 \prod_{i=1}^3  s_{*i}^{i\lambda_i}  ~.
\eeqn

Applying to MHV in \eqref{3ptMHVcelestialAmp}, we get the celestial super-amplitude  
\beqn
\mathscr M_3^{\text{MHV}}&=&
-\pi\delta(\sum_i \lambda_i) \frac{\delta(\bar z_{13}) \delta(\bar z_{23})  \sgn(z_{23}z_{31}) }{   z_{12}  z_{23}^2z_{31}^2  }\Theta\Big( \frac{\epsilon_1}{\epsilon_3} \frac{z_{23} }{  z_{12}} \Big)  \Theta\Big( \frac{ \varepsilon_2}{ \varepsilon_3} \frac{z_{31} }{  z_{12}} \Big)   \frac{z_{12}^4}{ s_{*1}s_{*2}s_{*3}^2  }
\nonumber \\& &
\times  \delta^4 \Big( \eta_1 e^{ \frac12 \p_{\Delta_1}} - \frac{z_{23}}{z_{12}}\eta_3 e^{ \frac12 \p_{\Delta_3}}  \Big) 
 \delta^4 \Big( \eta_2 e^{ \frac12 \p_{\Delta_2}}- \frac{z_{31}}{z_{12}}\eta_3  e^{ \frac12 \p_{\Delta_3}}\Big) 
 \prod_{i=1}^3  s_{*i}^{i\lambda_i}  
\\ &=&
-\pi\delta(\sum_i \lambda_i) \frac{\delta(\bar z_{13}) \delta(\bar z_{23})  \sgn(z_{23}z_{31}) }{   z_{12}  z_{23}^2z_{31}^2  }\Theta\Big( \frac{ \varepsilon_1}{ \varepsilon_3} \frac{z_{23} }{  z_{12}} \Big)  \Theta\Big( \frac{ \varepsilon_2}{ \varepsilon_3} \frac{z_{31} }{  z_{12}} \Big)  
\nonumber\\& &
\times  
  \prod_{A=1}^4 \sum_{\underset{i<j}{i,j=1}}^3 z_{ij} \sqrt{ \frac{s_{*i} }{s_{*3} }\frac{s_{*j} }{s_{*3} }   }    \eta_i^A\eta_j^A
\times   \Big( \frac{s_{*1}  }{s_{*3} } \Big)^{i\lambda_1-1}    \Big( \frac{s_{*2}  }{s_{*3} } \Big)^{i\lambda_2-1}  ~.
\label{3MHVsuper}
\eeqn
where we also present an equivalent expression in the second equality for later discussions. 

Picking up the term $\eta_1^4 \eta_2^4$, one gets
\beqn
\mathscr  M_3^{\text{MHV}}\Big|_{\eta_1^4 \eta_2^4}&=&
-\pi\delta(\sum \lambda_i) \frac{\delta(\bar z_{13}) \delta(\bar z_{23})  \sgn(z_{23}z_{31}) }{z_{12}  z_{23}^2z_{31}^2  }\Theta\Big( \frac{\varepsilon_1}{\varepsilon_3} \frac{z_{23} }{  z_{12}} \Big)  \Theta\Big( \frac{ \varepsilon_2}{ \varepsilon_3} \frac{z_{31} }{  z_{12}} \Big)   \frac{z_{12}^4}{ s_{*1}s_{*2}s_{*3}^2  }
   s_{*1}^{i\lambda_1+2}     s_{*2}^{i\lambda_2+2}     s_{*3}^{i\lambda_3}  
\nonumber\\&=&
-\pi  \delta(\sum \lambda_i)  \frac{z_{12}^3}{z_{23}^2z_{31}^2}
\sgn(z_{23}z_{31}) \delta(\bar z_{13})\delta(\bar z_{23}) 
\Theta\Big( \frac{ \varepsilon_1}{ \varepsilon_3} \frac{z_{23} }{  z_{12}} \Big)  \Theta\Big( \frac{ \varepsilon_2}{ \varepsilon_3} \frac{z_{31} }{  z_{12}} \Big) 
     \Big(\frac{ \varepsilon_1}{ \varepsilon_3} \frac{z_{23}}{z_{12}}\Big)^{1+i\lambda_1}\Big( \frac{ \varepsilon_2}{ \varepsilon_3}  \frac{z_{31}}{z_{12}}\Big)^{1+i\lambda_2} 
      ~.    \nonumber\\
\eeqn
This agrees with \eqref{3ptMHVcelestialAmp}. 
 
One can also consider the $\overline{\mathrm {MHV}}$ superamplitude:
\beqn
\mathscr A^{ \overline{ \mathrm{  MHV}}}_3(\omega_j,z_j,\bar z_j)
&=&\frac{ \delta^4(P) \delta^4\Big(\SV{12} \eta_3+\SV{23}\eta_1+\SV{31}\eta_2 \Big)}{\SV{12} \SV{23}\SV{3 1}}  
\\
&=& 2\frac{ \omega_1\omega_2\omega_3\;  }{ \bar z_{12}\bar z_{23}\bar z_{31}}   \delta^4(P) 
\prod_{A=1}^4 \Big(  \varepsilon_1  \varepsilon_2 \frac{\bar z_{12}}{\sqrt{\omega_3}} \eta_3^A+  \varepsilon_2  \varepsilon_3 \frac{\bar z_{23}}{\sqrt{\omega_1}}\eta_1^A+ \varepsilon_1  \varepsilon_3 \frac{\bar z_{31}}{\sqrt{\omega_2}} \eta_2^A \Big)~,
\eeqn
 %
%
whose celestial counterpart can be similarly derived 
\beqn\label{antiMHV3pt}
\mathscr M_3^{ \overline{ \mathrm{  MHV}}}&=&
 \pi\delta(\sum \lambda_i) \frac{\delta(  z_{13}) \delta(   z_{23})  \sgn(\bar z_{23}\bar z_{31}) }{    \bar z_{12}  \bar   z_{23}^2 \bar z_{31}^2  }\Theta\Big( \frac{ \varepsilon_1}{ \varepsilon_3} \frac{\bar z_{23} }{\bar  z_{12}} \Big)  \Theta\Big( \frac{ \varepsilon_2}{ \varepsilon_3} \frac{\bar z_{31} }{\bar  z_{12}} \Big)   { \bar  s_{*1}  \bar  s_{*2}  }
\nonumber\\& &
\times   \delta^4 \Big( \varepsilon_1 \varepsilon_2 \bar z_{12} \; \eta_3  \; e^{-\frac12 \p_{\Delta_3}}+
 \varepsilon_2  \varepsilon_3 \bar  z_{23} \; \eta_1  \; e^{-\frac12 \p_{\Delta_1}}
 + \varepsilon_1  \varepsilon_3 \bar z_{31} \; \eta_2  \; e^{-\frac12 \p_{\Delta_2}}
  \Big)
  \prod_{i=1}^3 \bar s_{*i}^{i\lambda_i}  ~,
\eeqn
where $\bar s_{*i}$ is given in \eqref{s123} by replacing $z\to \bar z$. Picking up the  term $\eta_3^4$, one recovers  the ${ \overline{ \mathrm{  MHV}}} $ for gluon three-point amplitude
\beqn
\mathscr M_3^{ \overline{ \mathrm{  MHV}}} |_{\eta_3^4 }&=&
 \pi\delta(\sum \lambda_i) \frac{  \bar z_{12} ^3 }{    \bar   z_{23}^2 \bar z_{31}^2  }
\delta(  z_{13}) \delta(   z_{23})  \sgn(\bar z_{23}\bar z_{31})
\Theta\Big( \frac{ \varepsilon_1}{ \varepsilon_3} \frac{\bar z_{23} }{\bar  z_{12}} \Big)  \Theta\Big( \frac{ \varepsilon_2}{ \varepsilon_3} \frac{\bar z_{31} }{\bar  z_{12}} \Big)   
 \Big(\frac{ \varepsilon_2}{ \varepsilon_3} \frac{\bar z_{31} }{\bar  z_{12}}  \Big)^{i\lambda_1+1}
  \Big(\frac{ \varepsilon_1}{ \varepsilon_3} \frac{\bar z_{23} }{\bar  z_{12}}  \Big)^{i\lambda_2+1}~.
  \nonumber\\
\eeqn

The full three-point celestial super-amplitude is the sum of two helicity configurations:     $\mathscr M_3=\mathscr M_3^{  { \mathrm{  MHV}}} +\mathscr M_3^{ \overline{ \mathrm{  MHV}}} $.
   
\section{Soft theorem and Ward identity}\label{softlimit}

In this section we study the soft limit of  $\mathcal N=4$ SYM theory.  The soft limit captures some universal properties of quantum field theory. Indeed it has been  understood in recent years that the soft theorems  can be regarded as the Ward identity of various  asymptotic symmetries. See \cite{Strominger:2017zoo} for a review.  For graviton, the soft theorems arise  from the BMS symmetries which include super-translation and super-rotation \cite{Strominger:2013jfa}. For  gluon, the underlying symmetry for the leading soft theorem is the large gauge transformation \cite{Strominger:2013lka}. However, the relations between symmetry and      soft theorem are obscure in the conventional momentum space where   the soft limit corresponds to taking    one particle soft with vanishing energy. 

Such a connection becomes much more obvious once we go to the celestial sphere and study the celestial amplitude.   It turns out in the celestial basis, the energetically soft theorem becomes the conformally soft theorem \cite{Fan:2019emx,Nandan:2019jas,Adamo:2019ipt,Pate:2019mfs}. Such a conformally soft theorem can be further translated  into the Ward identity in CCFT. The Ward identity is just the standard Ward identity  in CFT which relates the CFT correlators with and without the  insertion of the soft symmetry current. For  the leading soft gluon theorem, the corresponding soft symmetry current is just the Kac-Moody  current. 

We will generalize all these discussions to $\mathcal N=4$ SYM theory. We will show   the  conformally supersoft theorem  where one operator  has dimension $\Delta\to 1$. Translating into the language of CCFT, this corresponds to the super-Ward identity.

\subsection{Conformally soft  gluon theorem in  YM }
Let us review the conformally soft gluon theorem in YM theory first \cite{Pate:2019mfs}. 
In  the definition of celestial amplitude \eqref{MellinTsf}, one can choose the $j$-th particle and   take the limit $\lambda_j\to 0$ which leads to  
\be\label{softLmt}
\lim_{\lambda_j \to 0}   i \lambda_j   
 \mathcal M_n(\Delta_i,J_i,  z_i, \bar z_i)
 =2\Big(\prod_{  \underset  {k\neq j} {j=1 } }^n  \int_0^\infty d\omega_k\; \omega_k^{\Delta_k -1}  \Big)  
 \int_0^\infty d\omega_j \delta(\omega_j) \omega_j \mathcal A_n(\omega_i,z_i,\bar z_i )~.
\ee
It is easy to recgonize that the right hand side just picks the residue of momentum space amplitude $ \mathcal A_n(\omega_i,z_i,\bar z_i )$ at $\omega_j=0=p_j$.   The (energetically) soft  theorem for color-ordered partial amplitide is given by in \eqref{YMsoftlimit}, \eqref{softfactor}. In terms of the celestial coordinates, the leading soft  (positive helicity gluon) theorem takes the form: 
\be
\lim_{\omega_j\to 0} \omega_j \mathcal A_n(\omega_i,z_i,\bar z_i ) =-\frac12 \frac{z_{j-1,j+1}}{z_{j-1,j} z_{j,j+1}} \mathcal A_{n-1}(1,\cdots, j-1, j+1, \cdots,n )~,
\ee
where   the $j$-th particle is omitted on the right hand side.  Plugging this into \eqref{softLmt}, we get the tree-level conformally soft theorem \cite{Pate:2019mfs}
\be\label{confsofttheorem}
\lim_{\lambda_j \to 0}   i \lambda_j   \mathcal M_n(\Delta_i,J_i,  z_i, \bar z_i)
 =-\frac12 \frac{z_{j-1,j+1}}{z_{j-1,j} z_{j,j+1}} \mathcal M_{n-1}(1,\cdots, j-1, j+1, \cdots,n )~,
 \ee
where the factor can also be written as $1/z_{j-1,j}+1/z_{j,j+1}  $.

The conformal correlator  is related to the full amplitude  which includes the color factors and takes the following form at tree level:
\be\label{FullAmpCorrelator}
\EV{O_{\Delta_1,J_1} ^{ a_1}(z_1,\bar z_1)O_{\Delta_2,J_2} ^{ a_2}(z_2,\bar z_2)\cdots
O_{\Delta_n,J_n} ^{ a_n}(z_n,\bar z_n)}
=g_{\text{YM}}^{n-2}\sum_{\sigma\in S_{n-1}}  \mathcal M_n(1 ,\sigma_2 , \cdots, \sigma_n ) \Tr (T^{a_1}T^{a_{\sigma_2}} \cdots T^{a_{\sigma_n}}) ~,
\ee
where $\sigma$ acts by permuting the label $\{ 2, 3, \cdots,n\}$.

If we   define 
\be
J^a(z) =\lim_{\Delta\to 1} (\Delta -1) O_{\Delta ,+} ^{ a  }~,
\ee
which has conformal weights $(h ,\bar h)=(1,0)$, then the conformally soft theorem \eqref{confsofttheorem}  becomes the Ward identity  of a current algebra (after choosing normalization properly):  
\be\label{JOWI}
 \EV{J^a(z)O_{\Delta_1,J_1} ^{ b_1}(z_1,\bar z_1)\cdots
O_{\Delta_n,J_n} ^{ b_n}(z_n,\bar z_n)}
=\sum_{j=1}^n \frac{f^{ab_j c}}{z-z_j} \EV{ O_{\Delta_1,J_1} ^{ b_1}(z_1,\bar z_1)\cdots 
O_{\Delta_j,J_j} ^{  c}(z_j,\bar z_j)\cdots
O_{\Delta_n,J_n} ^{ b_n}(z_n,\bar z_n)}~,
\ee
where $f^{abc}$ is the structure constant of the Lie algebra.

In terms of OPE, we have
\be\label{JOope}
J^a(z)  O_{\Delta,J}^{  b}(w,\bar w)\sim \frac{f^{abc}}{z-w}O_{\Delta,J}^{  c}(w,\bar w)~.
\ee

We can also consider the soft     theorem for negative helicity gluon. Especially we have an anti-holomorphic current defined as
\be
\bar J^a(\bar z) =\lim_{\Delta \to 1} (\Delta-1) O_{\Delta,- } ^{ a  } ~,
\ee
which obeys similar Ward identity and OPE as that in \eqref{JOWI} \eqref{JOope}.

So far, we only considered the leading soft theorem, namely the first term in \eqref{softfactor}.  The second term in  \eqref{softfactor} corresponds to the subleading soft theorem.  After performing Mellin transformation, one obtains \cite{Pate:2019lpp,Banerjee:2020vnt,Fotopoulos:2020bqj}
\beqn\label{KOWI}
&&
 \EV{K^a(z,\bar z)O_{\Delta_1,J_1} ^{ b_1}(z_1,\bar z_1)\cdots
O_{\Delta_n,J_n} ^{ b_n}(z_n,\bar z_n)}
\nonumber\\&=&
\sum_{j=1}^n \frac{ \varepsilon_j\; f^{ab_j c}}{z-z_j}\Big(-2\bar h_j+1+(\bar z-\bar z_j) \frac{\p}{\p\bar z_j}\Big) \EV{ O_{\Delta_1,J_1} ^{ b_1}(z_1,\bar z_1)\cdots 
O_{\Delta_j-1,J_j} ^{ c}(z_j,\bar z_j)\cdots
O_{\Delta_n,J_n} ^{ b_n}(z_n,\bar z_n)}
\qquad
\eeqn
where 
\be
K^a(z,\bar z) =\lim_{\Delta\to 0}  \Delta O_{\Delta,+ } ^{ a  }~.
\ee

\subsection{Conformally supersoft    theorem in  SYM }
The (leading) super-soft theorem of SYM is given by  \eqref{SYMsofttheorem} and \eqref{supersoftfactor}. In terms of celestial coordinates, it reads
\be
\omega_j \mathscr A_n ( \cdots, j-1, j  , j+1, \cdots) \xrightarrow{\omega_j\to 0} 
-\frac12 \Big( \frac{z_{j-1,j+1}}{z_{j-1,j} z_{j,j+1}}   + \frac{\bar z_{j-1,j+1}}{\bar z_{j-1,j} \bar z_{j,j+1}}   \delta^4(\eta_j)      \Big)  \mathscr  A_n ( \cdots, j-1, j+1, \cdots) ~.
\ee 
The formula \eqref{softLmt} can also be used for super-amplitude  above and gives rise to 
\be\label{confsofttheorem}
\lim_{\lambda_j \to 0}   i \lambda_j   \mathscr M_n( \cdots, j-1, j  , j+1, \cdots)
 =-\frac12\Big( \frac{z_{j-1,j+1}}{z_{j-1,j} z_{j,j+1}}   + \frac{\bar z_{j-1,j+1}}{\bar z_{j-1,j} \bar z_{j,j+1}}   \delta^4(\eta_j)      \Big)  \mathscr M_{n-1}( \cdots, j-1,   j+1, \cdots)~.
\ee
This is just the leading conformally super-soft theorem in SYM.  In the following we will verify that this relation holds  for the four-point  celestial superamplitude.

The four-point celestial superamplitude  is computed in \eqref{4ptCamp2} and written in terms of kinematic factors and cross-ratio. For the soft theorem, it is more convenient to use the following  equivalent expression:
 \beqn
\mathscr M_4&=&	\frac{\pi}{2} (\sum_{i=1}^4\lambda_i) \delta \Big(z_{12} z_{34} \bz_{13} \bz_{24}-z_{13} z_{24} \bz_{12} \bz_{34} \Big)   \frac{1}{z_{12}z_{23}z_{34}z_{41}}	
\prod_{A=1}^4 \Big(\sum_{\underset{i<j}{i,j=1}}^4z_{ij}\sqrt{ \frac{s_{*i} }{s_{*3} }\frac{s_{*j} }{s_{*3} }   }   \eta_i^A \eta_j^A \Big)
\nonumber\\&& \times	
    \Big(\frac{s_{*1}}{s_{*3}}  \Big)^{-1+ i \lambda_1}
     \Big(\frac{s_{*2}}{s_{*3}} \Big)^{-1+i  \lambda_2}
 	\Big(\frac{s_{*4}}{s_{*3}} \Big)^{-1+i  \lambda_4} 
	  \Theta    \Big(\frac{s_{*1}}{s_{*3}}  \Big) 	
	    \Theta    \Big(\frac{s_{*2}}{s_{*3}}  \Big) 	  \Theta    \Big(\frac{s_{*4}}{s_{*3}}  \Big) ~,
  \eeqn
  where
  \be\label{s123ex}
\frac{s_{*1}}{s_{*3}} =\frac{ \varepsilon_3}{ \varepsilon_1}\frac{z_{23}\bar z_{34}}{z_{12}\bar z_{14}}
 =\frac{ \varepsilon_3}{ \varepsilon_1}\frac{ \bar z_{23}  z_{34}}{ \bar z_{12}  z_{14}}~, \quad
\frac{s_{*2}}{s_{*3}} =\frac{ \varepsilon_3}{ \varepsilon_2}\frac{z_{13}\bar z_{34}}{z_{12}\bar z_{42}}
 =\frac{ \varepsilon_3}{ \varepsilon_2}\frac{\bar z_{13}  z_{34}}{\bar z_{12}  z_{42}}~, \quad
\frac{s_{*4}}{s_{*3}} =\frac{ \varepsilon_3}{ \varepsilon_4}\frac{z_{23}\bar z_{13}}{z_{42}\bar z_{14}}
=\frac{ \varepsilon_3}{ \varepsilon_4}\frac{\bar z_{23}  z_{13}}{\bar z_{42}  z_{14}}~.
\ee
For each term, the equalities of the two different expressions arise  from the delta-function $ \delta (z_{12} z_{34} \bz_{13} \bz_{24}-z_{13} z_{24} \bz_{12} \bz_{34}  )$.~\footnote{Note that  since we want to relate this four-point amplitude to three-point  on-shell  amplitude which only makes sense  in split signature, we will thus also consider here the four-point amplitude in the split signature, namely here $z_{ij}$ and $\bar z_{ij}$ are all real and independent. 
 }
 
 Using the following identity \cite{Pate:2019mfs}
 \be
\lim_{\epsilon\to 0}\frac{\epsilon}{2} |x|^{\epsilon-1}=\delta(x)~,
\ee	
we get
   \beqn \label{4ptLimit}
\lim_{\lambda_4\to 0} i\lambda_4 \mathscr  M_4&=&	\frac{\pi}{2} (\sum_{i=1}^4\lambda_i) \delta \Big(z_{12} z_{34} \bz_{13} \bz_{24}-z_{13} z_{24} \bz_{12} \bz_{34} \Big)   \frac{1}{z_{12}z_{23}z_{34}z_{41}}	
\prod_{A=1}^4 \Big(\sum_{\underset{i<j}{i,j=1}}^4z_{ij} \sqrt{ \frac{s_{*i} }{s_{*3} }\frac{s_{*j} }{s_{*3} }   }  \eta_i^A \eta_j^A \Big)
\nonumber\\&&
 \times
     \Big(\frac{s_{*1}}{s_{*3}}  \Big)^{-1+ i \lambda_1}
     \Big(\frac{s_{*2}}{s_{*3}} \Big)^{-1+i  \lambda_2}
 \delta	\Big(\frac{s_{*4}}{s_{*3}} \Big)   
     	  \Theta    \Big(\frac{s_{*1}}{s_{*3}}  \Big) 	
	    \Theta    \Big(\frac{s_{*2}}{s_{*3}}  \Big) 	 ~,
  \eeqn
  where we used $\Theta(0)=1/2$.

 Examining the two expressions for $\frac{s_{*4}}{s_{*3}}$ in \eqref{s123ex}, we see that  $  \frac{s_{*4}}{s_{*3}} =0  $ has two branches $z_{13}=0$ or $\bar z_{13}=0$.  \footnote{Note that we are now in the split signature, so $z_{ij} $ and $\bar z_{ij}$ are independent real variables. Other choices, say $z_{23}=0$, give  vanishing or infinite $  {s_{*1}}/{s_{*3}} $  or $  {s_{*2}}/{s_{*3}} $.}
  
  In the first branch $\bar z_{13}= 0$, the delta functions can be simplified
\beqn \label{deltasimp1}
		\delta \Big(\frac{ \varepsilon_3}{ \varepsilon_4}\frac{\bz_{13} z_{23}  }{\bz_{14} z_{42}  } \Big) 
		&=& {\rm sgn}(\bz_{14} z_{42} z_{23}) \frac{\bz_{14} z_{42} }{ z_{23} } \delta (\bz_{13})~, 
 \\
\delta \Big(z_{12} z_{34} \bz_{13} \bz_{24}-z_{13} z_{24} \bz_{12} \bz_{34} \Big) &=&
		\delta ( z_{13} z_{24} \bz_{12} \bz_{34}  )   =\frac{ {\rm sgn}(z_{13} z_{24}   \bz_{34} ) }{z_{13} z_{24}   \bz_{34}} \delta( \bz_{12})~,\quad
	\\	 \rightarrow  
 \delta \Big(z_{12} z_{34} \bz_{13} \bz_{24}-z_{13} z_{24} \bz_{12} \bz_{34} \Big)	\delta \Big(\frac{ \varepsilon_3}{ \varepsilon_4}\frac{\bz_{13} z_{23}  }{\bz_{14} z_{42}  } \Big) 
&=&
 {\rm sgn}( z_{23} z_{31}   ) \frac {1  }{   z_{23}  z_{31}  } \delta( \bz_{12})  \delta (\bz_{13})	~,
 \eeqn
so in this branch we have $\bar z_1=\bar z_2=\bar z_3$ and 
%
%
   \be 
\frac{s_{*1}}{s_{*3}} =\frac{ \varepsilon_3}{ \varepsilon_1}\frac{z_{23}}{z_{12} }~, \quad
\frac{s_{*2}}{s_{*3}} =\frac{ \varepsilon_3}{ \varepsilon_2}\frac{z_{31} }{z_{12} }~. 
\ee
 
Substituting them back into \eqref{4ptLimit}, one finds	
   \beqn
 \lim_{\lambda_4\to 0} i\lambda_4\mathscr  M_4\Big|_{\bar z_{13}=0}
  &=&
	 \frac{ \pi}{2} (\sum_{i=1}^4\lambda_i)  \frac{1}{z_{12}z_{23}z_{34}z_{41}}
\prod_{A=1}^4 \Big(\sum_{\underset{i<j}{i,j=1}}^4 z_{ij} \sqrt{ \frac{s_{*i} }{s_{*3} }\frac{s_{*j} }{s_{*3} }   }   \eta_i^A \eta_j^A \Big)
     \Big(\frac{s_{*1}}{s_{*3}}  \Big)^{-1+ i \lambda_1}
     \Big(\frac{s_{*2}}{s_{*3}} \Big)^{-1+i  \lambda_2}
     \nonumber\\&&\times
 {\rm sgn}( z_{23} z_{31}   ) \frac {1  }{   z_{23}  z_{31}  } \delta( \bz_{12})  \delta (\bz_{13})		
     	  \Theta    \Big(\frac{s_{*1}}{s_{*3}}  \Big) 	
	    \Theta    \Big(\frac{s_{*2}}{s_{*3}}  \Big) 	 
  \\&=&
		 \frac{ \pi}{2}  (\sum_{i=1}^4\lambda_i) \frac {z_{31}  }{ z_{34}z_{41}      }    
		 \frac{ {\rm sgn}( z_{23} z_{31}   )  \delta( \bz_{12})  \delta (\bz_{13})	  }{z_{12}z_{23}^2  z_{31} ^2 }  
		     	  \Theta    \Big( \frac{ \varepsilon_3}{ \varepsilon_1}\frac{z_{23}}{z_{12} }  \Big) 	
	    \Theta    \Big(\frac{ \varepsilon_3}{ \varepsilon_2}\frac{z_{31} }{z_{12} }  \Big) 	 
     \nonumber\\&&\times
     	         \Big( \frac{ \varepsilon_3}{ \varepsilon_1}\frac{z_{23}}{z_{12} } \Big)^{-1+ i \lambda_1}
     \Big( \frac{ \varepsilon_3}{ \varepsilon_2}\frac{z_{31} }{z_{12} }\Big)^{-1+i  \lambda_2}
\prod_{A=1}^4 \Big(\sum_{\underset{i<j}{i,j=1}}^3 z_{ij} \sqrt{ \frac{s_{*i} }{s_{*3} }\frac{s_{*j} }{s_{*3} }   }   \eta_i^A \eta_j^A \Big)
  \eeqn	
  where in the parentheses   $i,j$ are summed over 1 to 3 because $s_{*4}=0$  and has no contribution.

Compared to \eqref{3MHVsuper}, we find 
\be\label{limzb}
\lim_{\lambda_4\to 0} i\lambda_4\mathscr   M_4\Big|_{\bar z_{13}=0} =-\frac12 \frac{z_{31} }{z_{34}z_{41}} \mathscr M_3^{\mathrm{MHV}}~.
\ee

In the other branch $z_{13}=0$, similarly  we have 
   \beqn\label{M4z13}
\lim_{\lambda_4\to 0} i\lambda_4 \mathscr M_4\Big|_{  z_{13}=0}
&=&
	 \frac{ \pi}{2} (\sum_{i=1}^4\lambda_i)  
\prod_{A=1}^4 \Big(\sum_{\underset{i<j}{i,j=1}}^4 z_{ij} \sqrt{ \frac{s_{*i} }{s_{*3} }\frac{s_{*j} }{s_{*3} }   }   \eta_i^A \eta_j^A \Big)
     \Big(\frac{s_{*1}}{s_{*3}}  \Big)^{-1+ i \lambda_1}
     \Big(\frac{s_{*2}}{s_{*3}} \Big)^{-1+i  \lambda_2}
     \nonumber\\&&\times
 {\rm sgn}( \bar z_{23} \bar z_{31}   ) \frac {1  }{   \bar z_{23} \bar z_{31}  } 
   \frac{1}{z_{12}z_{23}z_{34}z_{41}}
 \delta( z_{12}) \delta( z_{13})
     	  \Theta    \Big(\frac{s_{*1}}{s_{*3}}  \Big) 	
	    \Theta    \Big(\frac{s_{*2}}{s_{*3}}  \Big) 	 ~.
  \eeqn		
The result seems to be divergent due to the factor $1/(z_{12}z_{23})$ and $\delta(z_{13})\delta(z_{12})$.	However the coefficient  $ z_{ij} \sqrt{s_{*i} s_{*j}}$  in the  Grassmann term also vanishes because $s_{*4}=0$ and $z_{ij}=0$ for $i,j\in \{1,2,3\}$. These two factors may cancel and lead  to a finite result. 
Indeed,  we have $z_{ij} \propto z_{12}$ and $\sqrt{s_{*4}} \propto \sqrt{z_{12}}$. So in order to cancel the factor $z_{12}^2$, the only possibility is to take  $z_{i4} \sqrt{s_{*i} s_{*4}}$ for each $A$. Other choices would lead to higher order zeros and thus vanishing contribution. 
 
Especially, we find 
\be
\frac{z_{i4} \sqrt{ \frac{s_{*i} }{s_{*3} }\frac{s_{*4} }{s_{*3} }   }  }{ \varepsilon_j \varepsilon_k \; \bar z_{jk}/\sqrt{s_{*i}}}= 
 \varepsilon_1 \varepsilon_2 \sqrt{ \varepsilon_3  \varepsilon_4}
 \sqrt{s_{*3}  }  \sqrt{\frac{z_{12}z_{14}\bar z_{13}\bz_{32}}{\bz_{12}^3\bz_{34}}} , \qquad
(i,j,k)=(1,2,3),(2,3,1),(3,1,2)~,
\ee
which enables us to rewrite
\beqn
\prod_{A=1}^4 \Big(\sum_{i }^3 z_{i4} \sqrt{ \frac{s_{*i} }{s_{*3} }\frac{s_{*4} }{s_{*3} }   }   \eta_i^A \eta_4 ^A \Big)
&=&
 \delta^4\Big(\sum_{i=1 }^3 \varepsilon_j \varepsilon_k \; \bar z_{jk}/\sqrt{s_{*i}}   \eta_i  \Big) \delta^4(\eta_4) \;
 s_{*3}^2   \; \Big(\frac{z_{12}z_{14}\bar z_{13}\bz_{32}}{\bz_{12}^3\bz_{34}}\Big)^2~.
\eeqn
With this identity, \eqref{M4z13} can be rewritten as
\beqn
\lim_{\lambda_4\to 0} i\lambda_4 \mathscr M_4\Big|_{  z_{13}=0}&=&
	 \frac{ \pi}{2} (\sum_{i=1}^4\lambda_i)  
 \delta^4\Big( \frac{  \varepsilon_2 \varepsilon_3\bar  z_{23}}{\sqrt{s_{*1}}  } \eta_1  +  \frac{ \varepsilon_1 \varepsilon_3 \bar z_{31} }{\sqrt{s_{*2}}  }  \eta_2  + \frac{  \varepsilon_1 \varepsilon_2\bar z_{12}}{\sqrt{s_{*3}}  }  \eta_3   \Big) \delta^4(\eta_4) 
 {\rm sgn}( \bar z_{23} \bar z_{31}   )    \delta( z_{12}) \delta( z_{13})
     \nonumber\\&&\times
      s_{*3}^2    \Big(\frac{z_{12}z_{14}\bar z_{13}\bz_{32}}{\bz_{12}^3\bz_{34}}\Big)^2
 \frac {1  }{   \bar z_{23} \bar z_{31}  } 
   \frac{1}{z_{12}z_{23}z_{34}z_{41}}
     \Big(\frac{s_{*1}}{s_{*3}}  \Big)^{-1+ i \lambda_1}
     \Big(\frac{s_{*2}}{s_{*3}} \Big)^{-1+i  \lambda_2}
          	  \Theta    \Big(\frac{s_{*1}}{s_{*3}}  \Big) 	
	    \Theta    \Big(\frac{s_{*2}}{s_{*3}}  \Big) 	 
\nonumber   \\&=&
- \frac{ \pi}{2} (\sum_{i=1}^4\lambda_i)  
 \delta^4\Big( \frac{ \varepsilon_2 \varepsilon_3\bar  z_{23}}{\sqrt{s_{*1}}  } \eta_1  +  \frac{ \varepsilon_1 \varepsilon_3 \bar z_{31} }{\sqrt{s_{*2}}  }  \eta_2  + \frac{  \varepsilon_1 \varepsilon_2\bar z_{12}}{\sqrt{s_{*3}}  }  \eta_3   \Big) \delta^4(\eta_4) 
 {\rm sgn}( \bar z_{23} \bar z_{31}   )    \delta( z_{12}) \delta( z_{13}) 
      \nonumber\\&&\times
      s_{*3}^2 \;
      \frac{\bz_{31}}{\bz_{34}\bz_{41}} \frac{1}{\bz_{12}\bz_{23}^2\bz_{31}^2}
     \Big(\frac{s_{*1}}{s_{*3}}  \Big)^{ 1+ i \lambda_1}
     \Big(\frac{s_{*2}}{s_{*3}} \Big)^{ 1+i  \lambda_2}
  	  \Theta    \Big(\frac{s_{*1}}{s_{*3}}  \Big) 	
	    \Theta    \Big(\frac{s_{*2}}{s_{*3}}  \Big) 	 ~,
\eeqn
where we used the following relation (see \eqref{s123ex})
  \be 
\frac{s_{*1}}{s_{*3}}    =\frac{ \varepsilon_3}{ \varepsilon_1}\frac{ \bar z_{23}  }{ \bar z_{12}  }~, \quad
\frac{s_{*2}}{s_{*3}}  =\frac{ \varepsilon_3}{ \varepsilon_2}\frac{\bar z_{31}  }{\bar z_{12}  }~, \quad
\frac{z_{23}}{z_{12}} =\frac{\bz_{23}\bz_{14}}{\bz_{12}\bz_{34}}~.
 \ee 
 Compared to \eqref{antiMHV3pt}, we find 
 \be\label{limz}
 \lim_{\lambda_4\to 0} i\lambda_4 \mathscr M_4\Big|_{  z_{13}=0}=  -\frac12  \delta^4(\eta_4)   \frac{\bz_{31}}{\bz_{34}\bz_{41}} 
 \mathscr M_3^{ \overline{ \mathrm{  MHV}}}~.
 \ee
 Combining \eqref{limzb} and \eqref{limz} together, we obtain the conformally supersoft theorem 
  \be 
 \lim_{\lambda_4\to 0} i\lambda_4 \mathscr M_4 =    -\frac12  
 \Big(  \frac{ z_{31}}{ z_{34} z_{41}}  \mathscr M_3^{  { \mathrm{  MHV}}}+\delta^4(\eta_4)   \frac{\bz_{31}}{\bz_{34}\bz_{41}}  \mathscr M_3^{ \overline{ \mathrm{  MHV}}} \Big) 
 = -\frac12 
\Big(  \frac{ z_{31}}{ z_{34} z_{41}} +\delta^4(\eta_4)   \frac{\bz_{31}}{\bz_{34}\bz_{41}}  \Big) 
 \mathscr M_3 ~,
 \ee
 where in  the second equality   the cross-terms disappear as    $z_{13}\delta(z_{13})=\bar z_{13} \delta(\bar z_{13})=0$. This thus gives the first non-trivial explicit check of the conformally supersoft theorem.

\subsection{Super-Ward identity from conformally supersoft theorem}

Following the discussion  for pure YM theory, we can recast the  conformally supersoft theorem  \eqref{confsofttheorem}  in terms of the conformal correlators, leading to the super-Ward identity in CCFT
 \beqn
 &&\EV{\cJ^a(z,\bar z,\eta)\cO_{\Delta_1} ^{h_1b_1}(z_1,\bar z_1)\cdots
\cO_{\Delta_n,J_n} ^{ b_n}(z_n,\bar z_n)}
\\ &=&\sum_{j=1}^n  \Big( \frac{f^{ab_j c}}{z-z_j} + \frac{f^{ab_j c}}{\bar z- \bar z_j}   \delta^4(\eta) \Big)\EV{ \cO_{\Delta_1,J_1} ^{ b_1}(z_1,\bar z_1)\cdots 
\cO_{\Delta_j,J_j} ^{  c}(z_j,\bar z_j)\cdots
\cO_{\Delta_n,J_n} ^{ b_n}(z_n,\bar z_n)}
\eeqn
Then we can read the  following  OPE  
 \be\label{JOsupef}
\cJ^a( z_1, \bar z_1,\eta_1) \cO_{\Delta    }^b(z_2, \bar z_2,\eta_2)
\sim
  \frac{f^{abc}}{z_{12}}  \cO^c_{\Delta }(z_2, \bar z_2,\eta_2)   \nonumber
 +
 \frac{f^{abc}}{\bar z_{12} } \delta^4(\eta_1)  \cO^c_{\Delta }(z_2, \bar z_2, \eta_2)    
\ee 
where we introduced  
\be 
\cJ^a(z, \bar z,\eta) =\lim_{\Delta\to 1}  (\Delta-1)\cO_\Delta (z, \bar z,\eta) ~,
\ee  
and $\mathcal O_\Delta$ is just the superoperator (essentially the celestial on-shell superfield \eqref{celestialSuperfield})
\beqn\label{ }
\mathcal O_{\Delta}(z,\bar z,   \eta)&=&\mathsf G_\Delta^+(z,\bar z )+\eta^A \mathsf\Gamma_{A\; \Delta}(z,\bar z )
+\frac{1}{2!}\eta^A\eta^B \mathsf\Phi_{AB\; \Delta}(z,\bar z  )
\\&&
+\frac{1}{3!}\epsilon_{ABCD} \eta^A\eta^B \eta^C \bar{\mathsf\Gamma}^{D}_\Delta(z,\bar z )
+\frac{1}{4!}\epsilon_{ABCD} \eta^A\eta^B \eta^C\eta^D \mathsf G_\Delta^- (z,\bar z )~. 
\eeqn

The discussions so far are at leading order.
 In principle, we can also generalize to subleading order.   We leave the generalization to the future. 
 

\section{Collinear limit and OPE} \label{collimit}

Another useful property in CFT is the OPE: when two local operators in CFT are inserted at nearby points, this is equivalent to inserting a set of operators and superposing them  at the same point.  The spectra of operators and 
OPE coefficients are the defining datum of CFT. So to understand CCFT, we need to determine the OPE coefficients among all the celestial operators. 

It turns out that the OPE in CCFT can be naturally obtained from the collinear limit in the scattering amplitude. Indeed, in the collinear limit, the two particles travel along   approximately  the same direction and pass the nearby points in the celestial sphere, which is just the OPE limit. The translation from collinear limit to OPE is  achieved through   Mellin transformation.  For gluon and graviton, this has been done in  \cite{Fan:2019emx}. Actually one can also use the consistency of CCFT to bootstrap the OPE \cite{Pate:2019lpp}. 

We will generalize these results to the $\mathcal N=4$ SYM theory and obtain the super-OPE between super-operators via a Mellin transformation. We will also consider the OPE when    the conformal dimension of one of the operators  takes   special value  $\Delta=1,1/2,0, \cdots$, which correspond to conformally soft modes. At these values, the OPE has a pole and  the operator involved  is the soft ``super''-current  of the CCFT.

\subsection{OPE in YM}
The YM amplitude in the collinear limit is given in \eqref{YMcollinear} - \eqref{YMsplitfactor}. Written in terms of celestial variables, they read ($\omega=\omega_1+\omega_2$) \footnote{For concreteness, we assume that the collinear  particles are all out-going. Generalizations to case with in-coming particles are straightforward.}
\beqn\label{APPcolliear}
\mathcal A_n (1^+, 2^+, \cdots ) &\xrightarrow{p_1\slash\!\slash p_2} &
   \frac{\omega }{\omega_1\omega_2} \frac{1}{ z_{12} }   \mathcal A_{n-1} (P_{12}^+\; , \cdots )  ~,
\\ \label{AMPcolliear}
\mathcal A_n (1^-, 2^-, \cdots ) &\xrightarrow{p_1\slash\!\slash p_2} &
   \frac{\omega }{\omega_1\omega_2} \frac{1}{ \bar z_{12} }  \mathcal  A_{n-1} (P_{12}^-\; , \cdots )  ~,
\\ \label{AMMcolliear}
\mathcal A_n (1^-, 2^+, \cdots ) &\xrightarrow{p_1\slash\!\slash p_2} &
   \frac{\omega_1 }{\omega \omega_2} \frac{1}{ z_{12} }  \mathcal  A_{n-1} (P_{12}^-\; , \cdots )  
   +
      \frac{\omega_2 }{\omega \omega_1} \frac{1}{  \bar z_{12} }  \mathcal  A_{n-1} (P_{12}^+\; , \cdots )  ~.
\eeqn
We would like to perform a Mellin transformation in order to obtain an equation  in terms  of conformal correlators. 
 For this purpose, let use first consider the Mellin transformation of  $\omega_1^\alpha \omega_2^\beta (\omega_1+\omega_2)^\gamma$ in the split factor:
\footnote{We used $
B(x,y) =\int_0^1 dt\; t^{x-1} (1-t)^{y-1} =\frac{\Gamma(x) \Gamma(y)}{\Gamma(x+y)}
$.}
\beqn\label{collinearBfcn}
&& \int_0^\infty \!\!   d\omega_2\;  \omega_2^{\Delta_2-1} \int_0^\infty  \!\!    d\omega_1\; \omega_1^{\Delta_1-1} 
\omega_1^\alpha \omega_2^\beta (\omega_1+\omega_2)^\gamma \; f(\omega_1+\omega_2)
= B\Big(\Delta_1+\alpha,\Delta_2+\beta \Big)    \!\!   
  \int_0^\infty d\omega \; \omega^{ \Delta_P  -1}  \; f(\omega)~,
  \qquad\nonumber\\
\eeqn
where  $\Delta_P=\Delta_1+\Delta_2+\alpha+\beta+\gamma$.

Using this formula and bearing in mind the identification of correlator with full amplitude \eqref{FullAmpCorrelator}, we can rewrite \eqref{APPcolliear},\eqref{AMPcolliear},\eqref{AMMcolliear} in terms of correlators and then extract the following OPEs  \cite{Fan:2019emx,Pate:2019lpp}:
\beqn\label{OPEPP}
O_{\Delta_1,+ }^{a}(z_1, \bar z_1) O_{\Delta_2 ,+ }^{b}(z_2, \bar z_2)  &\sim&
 \frac{f^{abc}}{z_{12}}   B\Big(\Delta_1-1,\Delta_2-1 \Big) O^{ c}_{\Delta_1+\Delta_2-1,+}(z_2, \bar z_2) ~,
\\ \label{OPEmm}
O_{\Delta_1,- }^{ a}(z_1, \bar z_1) O_{\Delta_2 ,- }^{b}(z_2, \bar z_2) &\sim&
 \frac{f^{abc}}{\bar z_{12}}   B\Big(\Delta_1-1,\Delta_2-1 \Big) O^{c}_{\Delta_1+\Delta_2-1,-}(z_2, \bar z_2)~, 
 \\ \label{OPEmp}
O_{\Delta_1,+ }^{a}(z_1, \bar z_1) O_{\Delta_2 ,- }^{b}(z_2, \bar z_2) &\sim&
 \frac{f^{abc}}{z_{12}}   B\Big(\Delta_1-1,\Delta_2+1  \Big) O^{ c}_{\Delta_1+\Delta_2-1,-}(z_2, \bar z_2) 
\nonumber\\&&+ 
\frac{f^{abc}}{\bar z_{12}}   B\Big(\Delta_1+1,\Delta_2-1  \Big) O^{c}_{\Delta_1+\Delta_2-1,+}(z_2, \bar z_2)~.
 \eeqn
When $\Delta=1,0$, the operator $O_{\Delta} $ becomes  become  soft currents, and we recover the  OPE  involving soft current, namely \eqref{JOope} and the singular term in \eqref{KOWI}.

\subsection{OPE in SYM}

Now we switch to the OPE of super-operators in SYM theory by considering the Mellin transformation of  the collinear limit in  SYM theory given in \eqref{SYMcollinear}, \eqref{SYMsplit1},\eqref{SYMsplit2}.  We  note that the super-split factor \eqref{SYMsplit2} is the same as the split factor in \eqref{splitYM}, up to the dressing factors involving $\eta$'s. So we expect that \eqref{SYMsplit2}  will give rise to OPE similar to \eqref{OPEmm}, up to  the $\eta$-dressing factors. The dressing factors contain    $\sqrt{\sfz}=\sqrt{p_1/(p_1+p_2)}$ and $\sqrt{1-\sfz}=\sqrt{p_2/(p_1+p_2)}$. For collinear $p_1,p_2$, they can also be written as  $\sqrt{\sfz}=\sqrt{\omega_1/(\omega_1+\omega_2)}$ and $\sqrt{1-\sfz}=\sqrt{\omega_2/(\omega_1+\omega_2)}$. So an insertion of $\sqrt{\sfz}$ corresponds to shifting $\alpha\to \alpha+\frac12, \gamma\to \gamma-\frac12$ in \eqref{collinearBfcn}. 
This amounts to the action of $e^{\frac12 \p_{\Delta_1}}$ on $B(\Delta_1+\alpha,\Delta_2+\beta)$ in  \eqref{collinearBfcn}. Similarly $\sqrt{1-\sfz}$  amounts to the action of $e^{\frac12 \p_{\Delta_2}}$ on $B(\Delta_1+\alpha,\Delta_2+\beta)$ in  \eqref{collinearBfcn}. 

As a consequence, we obtain the following OPE
\beqn\label{superOPE}
&&
\cO_{\Delta_1 }^a(z_1, \bar z_1,\eta_1) \cO_{\Delta_2  }^b(z_2, \bar z_2,\eta_2)  \nonumber
\\&  \sim&
 \frac{f^{abc}}{z_{12}}   \int d^4\eta_3  \Bigg[  \prod_{A=1}^4 \Big(\eta_3^A -e^{ \frac12 \p_{\Delta_1}}\eta_1^A -e^{ \frac12 \p_{\Delta_2}}\eta_2^A \Big)  B\Big(\Delta_1-1,\Delta_2-1 \Big) \Bigg] \cO^c_{\Delta_1+\Delta_2-1}(z_1, \bar z_1,\eta_3)   \nonumber
 \\&   &+
 \frac{f^{abc}}{\bar z_{12}}  \int d^4\eta_3\Bigg[  \prod_{A=1}^4   \Big(\eta_1^A \eta_2^A  -
  e^{ \frac12 \p_{\Delta_2}}  \eta_1^A \eta_3^A + e^{ \frac12 \p_{\Delta_1}} \eta_2^A \eta_3^A\Big) B\Big(\Delta_1-1,\Delta_2-1 \Big)  \Bigg]\cO^c_{\Delta_1+\Delta_2-1}(z_1, \bar z_1,\eta_3)~.   \nonumber\\
\eeqn

This expression can be simplified. 
First of all, we notice  that  
\beqn
 \int d\eta (\eta-\theta) f(\eta)
=f(\theta)~.
\eeqn
This can be further generalized to 
\beqn
\int d^4\eta_3 \prod_{A=1}^4 (\eta_3^A -\eta^A) f(\eta_3)  =f(\eta)~.
\eeqn

Secondly, one can also show   
\be
\int d\eta_3 (\eta_1\eta_2 -\eta_1 \eta_3+\eta_2\eta_3) f(\eta_3)
 =(\eta_1-\eta_2) f(\frac{\eta_1+\eta_2}{2})
  =(\eta_1-\eta_2) f(\eta_2)~,
\ee
 as well as its generalization
\beqn
&&
\int d\eta_3^A  \Big(\eta_1^A \eta_2^A  -
  e^{ \frac12 \p_{\Delta_2}}  \eta_1^A \eta_3^A + e^{ \frac12 \p_{\Delta_1}} \eta_2^A \eta_3^A\Big)   f(\eta_3^A)
 =  
\Big(\eta_1^A e^{ \frac12 \p_{\Delta_2}}   -\eta_2^A  e^{  \frac12 \p_{\Delta_1}}   \Big)
 f\Big(\eta_2^A  e^{-  \frac12 \p_{\Delta_2}}  \Big)~.
 \qquad
\eeqn

With these identities, the OPE in \eqref{superOPE} can be rewritten as
\beqn  \label{supeOPEsimple}
&&
\cO_{\Delta_1 }^a(z_1, \bar z_1,\eta_1) \cO_{\Delta_2  }^b(z_2, \bar z_2,\eta_2)  \nonumber
\\&  \sim&
 \frac{f^{abc}}{z_{12}}     \cO^c_{\Delta_1+\Delta_2-1}\Big(z_2, \bar z_2, \eta_1   e^{ \frac12 \p_{\Delta_1}}+ \eta_2   e^{ \frac12 \p_{\Delta_2}} \Big)  
 B\Big(\Delta_1-1,\Delta_2-1 \Big) \nonumber
 \\&   &+
 \frac{f^{abc}}{\bar z_{12}}  
    \cO^c_{\Delta_1+\Delta_2-1}\Big(z_2, \bar z_2, \eta_2 e^{-  \frac12 \p_{\Delta_2}}  \Big) 
     \delta^4\Big( \eta_1   e^{ \frac12 \p_{\Delta_2}}- \eta_2   e^{ \frac12 \p_{\Delta_1}} \Big)  
 B\Big(\Delta_1-1 , \Delta_2-1 \Big)~.
\eeqn
Note that the dimension shifting operator $e^{\pm \frac12 \p_{\Delta_{1,2}}}$ only acts on the function $B(\Delta_1-1 , \Delta_2-1)$.  Also the holomorphic OPE singularity in the first term only appears when the total number of $\eta$'s are less than or equal to 4, while the anti-holomorphic  OPE singularity in the second term   appears when the total number of $\eta$'s are greater than or equal to  4.

  One can then perform the $\eta$ expansion on both sides and get the OPE of the component operators. 
  For example, when considering the lowest component  without $\eta_1,\eta_2$, one gets the OPE of two gluon operator with positive helicity \eqref{OPEPP}: 
\be
\sfG_{\Delta_1  }^{+a}(z_1, \bar z_1) \sfG_{\Delta_2  }^{+b}(z_2, \bar z_2) \sim
 \frac{  f^{ab c}}{z_{12}}   B\Big(\Delta_1-1,\Delta_2-1 \Big) \sfG^{ +c}_{\Delta_1+\Delta_2-1 }(z_2, \bar z_2) ~.
\ee
And the OPE of two gluon operator with opposite helicity can be extracted from the $\eta^4_2$ term:
\beqn
\sfG_{\Delta_1  }^{+a} (z_1, \bar z_1 ) \sfG_{\Delta_2  }^{ -b}(z_2, \bar z_2 ) 
&\sim&  \frac{f^{abc}}{z_{12}}    B (\Delta_1-1,\Delta_2+1  )  \sfG^{-c}_{\Delta_1+\Delta_2-1} (z_2, \bar z_2)
\nonumber\\&&
+ \frac{f^{abc}}{\bar z_{12}}    B (\Delta_1+1,\Delta_2-1  )  \sfG^{+c}_{\Delta_1+\Delta_2-1} (z_2, \bar z_2)
\eeqn
which agrees with \eqref{OPEmp}. 

More OPEs are as follows:
\beqn
\sfGamma_{A\; \Delta_1  }^{ a}(z_1, \bar z_1) \sfGamma_{B\; \Delta_2  }^{ b}(z_2, \bar z_2) & \sim&
 \frac{  f^{ab c}}{z_{12}}   B\Big(\Delta_1-\frac12,\Delta_2-\frac12 \Big) \sfPhi^{ c}_{AB\;\Delta_1+\Delta_2-1 }(z_2, \bar z_2) ~,
\\ 
\bar\sfGamma _{  \Delta_1  }^{ A\; a}(z_1, \bar z_1) \sfGamma_{B\; \Delta_2  }^{ b}(z_2, \bar z_2)
& \sim&
\delta^A_B  \frac{f^{abc}}{  z_{12}}   B\Big(\Delta_1+\frac12 ,\Delta_2- \frac12 \Big)  \sfG^{-c}_{\Delta_1+\Delta_2-1} (z_2, \bar z_2)
\nonumber\\&&
+ \delta^A_B   \frac{f^{abc}}{ \bar z_{12}}   B\Big(\Delta_1-\frac12 ,\Delta_2+\frac12 \Big)  \sfG^{+c}_{\Delta_1+\Delta_2-1} (z_2, \bar z_2)~,
\\
\sfGamma_{A\; \Delta_1  }^{ a}(z_1, \bar z_1) \sfG_{\Delta_2 }^{+ b}(z_2, \bar z_2) &\sim &
 \frac{f^{ab c}}{z_{12}}   B\Big(\Delta_1-\frac12 ,\Delta_2-1  \Big) \sfGamma ^{ c}_{A\; \Delta_1+\Delta_2-1 }(z_2, \bar z_2) ~,
\\
\sfGamma_{A\; \Delta_1  }^{ a}(z_1, \bar z_1) \sfG_{\Delta_2 }^{- b}(z_2, \bar z_2) &\sim &
 \frac{f^{ab c}}{\bar z_{12}}   B\Big(\Delta_1+\frac12 ,\Delta_2-1  \Big) \sfGamma ^{ c}_{A\; \Delta_1+\Delta_2-1 }(z_2, \bar z_2) ~,
\\
\sfPhi_{AB\; \Delta_1  }^{  a}(z_1, \bar z_1 )  \sfPhi_{CD\; \Delta_2  }^{  b}(z_2, \bar z_2 ) 
&\sim &
 \frac{f^{abc}}{z_{12}}   \epsilon_{ABCD} B (\Delta_1 ,\Delta_2   )\sfG^{-c}_{ \Delta_1+ \Delta_2-1} (z_2, \bar z_2)
\nonumber \\&&
+ \frac{f^{abc}}{\bar z_{12}}   \epsilon_{ABCD} B (\Delta_1 ,\Delta_2   )\sfG^{+c}_{ \Delta_1+ \Delta_2-1} (z_2, \bar z_2 )~.
\eeqn

%
%
%

 These OPEs   agree  with~\cite{Fotopoulos:2020bqj} up to the R-symmetry index $A$ and the scalar $\sfPhi$  which are absent in $\mathcal N=1$ super-Yang-Mills theory. 
All the rest OPEs can also be obtained but will not be written down explicitly here. So the super-OPE \eqref{supeOPEsimple} compactly includes all the component OPE. This shows the power of superspace.

\subsection{OPE of soft mode}
With super-OPE \eqref{supeOPEsimple}, we can now discuss further the OPEs involving the conformally soft modes by  setting the conformal dimension to special values. 

\subsubsection{$\Delta\to 1$}
We  can take the limit $\Delta_1\to 1$ and define 
\be\label{cJop}
\cJ(z, \bar z,\eta) =\lim_{\Delta\to 1}  (\Delta-1)\cO_\Delta (z, \bar z,\eta) ~.
\ee
As we will see, this corresponds to  conformally soft gluons of both helicities. The OPE of $\cJ$ with $\cO_{\Delta}$  can be obtained from  \eqref{supeOPEsimple}.  Especially, we only need to focus on the singular term in the limit $\Delta_1 \to 1$ on the right hand side of  \eqref{supeOPEsimple}.  This happens only when there is no $e^{\frac12\p_{\Delta_1}}$ acting on    $B(\Delta_ 1-1,\Delta_2-1)$. Moreover, one has 
\footnote{This is generically true unless $\Delta_2=1,0,-1,-2,\cdots$.}
\be\label{Bto1}
\lim_{\Delta_1 \to 1} B(\Delta_ 1-1,\Delta_2-1) =\frac{1}{\Delta_ 1-1}~.
\ee
Since this is independent of the value of $\Delta_2$, we can also drop the shifting $e^{\pm\frac12\p_{\Delta_2}}$ in   \eqref{supeOPEsimple}.  

As a consequence, one has 
\be\label{OPEJO}
\cJ^a( z_1, \bar z_1,\eta_1) \cO_{\Delta    }^b(z_2, \bar z_2,\eta_2)
\sim
  \frac{f^{abc}}{z_{12}}  \cO^c_{\Delta }(z_2, \bar z_2,\eta_2)    
 +
 \frac{f^{abc}}{\bar z_{12} } \delta^4(\eta_1)  \cO^c_{\Delta }(z_2, \bar z_2, \eta_2)~.  
\ee 
This is the same as \eqref{JOsupef} which was obtained based on the soft theorem. 

Since we only have $\eta_1^0$ and $\eta_1^4$ terms on the right-hand side, $\cJ$ just corresponds to the soft gluons. And we can indeed rewrite $\cJ$ in \eqref{cJop} as 
\be
\cJ^a(z, \bar z,\eta) =J^a(z) +\bar J^a(\bar z) \delta^4(\eta)  ~,
\ee
where
\be
 J^a(z) =\lim_{\Delta\to 1}  (\Delta-1)\sfG^{+a}_\Delta (z, \bar z,\eta)  , \qquad
\bar  J^a(\bar z) =\lim_{\Delta\to 1}  (\Delta-1)\sfG^{-a}_\Delta (z, \bar z,\eta)  ~.
\ee
The conformal weights  $(h,\bar h)$  of $J$ and $\bar J$ are   $(1,0)$ and $(0,1)$, respectively. 

In \eqref{OPEJO}, one can further  take the limit $\Delta\to 1$. This leads to 
\be
\cJ^a(z_1, \bar z_1,\eta_1) \cJ^b(z_2, \bar z_2,\eta_2)
\sim
 \frac{f^{abc}}{z_{12}}     \cJ^c  (z_2, \bar z_2,  \eta_2     )  
 +
 \frac{f^{abc}}{\bar z_{12}}  
    \cJ^c (z_2, \bar z_2,\eta_2 ) 
     \delta^4 ( \eta_1  )~.
     \ee
Note that the right-hand side is not symmetric under the exchange of 1,2.  This just reflects a subtlety here: the OPE of two   gluons  with opposite helicity depends on the order of soft limits. Mathematically the subtlety arises because  \eqref{Bto1} breaks down if $\Delta_2=1$.

 \subsubsection{$\Delta\to \frac12$}
 
We can also take   the limit $\Delta \to 1/2$ and define  
 \be 
\mathcal K ^a(z, \bar z,\eta) =\lim_{\Delta \to \frac12}(\Delta -\frac12) \cO_{\Delta  }^a(z, \bar  z ,\eta )  ~.
\ee
 
 Following similar arguments above, here we allow exactly one  $e^{\frac12\p_{\Delta_1}}$ acting on    $B(\Delta_ 1-1,\Delta_2-1)$ in \eqref{supeOPEsimple}. This leads to the coefficient $B(\Delta_ 1-1/2,\Delta_2-1)=1/(\Delta_1-1/2)$ in the limit $\Delta_1\to 1/2$.

The   OPE \eqref{supeOPEsimple} then becomes
\be
\mathcal K ^a(z_1, \bar z_1,\eta_1) \cO_{\Delta   }^b(z_2, \bar z_2,\eta_2)
\sim
 \frac{f^{abc}}{z_{12}}     \cO^c_{\Delta  -\frac12}\Big(z_2, \bar z_2, \eta_1  + \eta_2     \Big)  
\Big|_{\eta_1 }
+
 \frac{f^{abc}}{\bar z_{12}}       \delta^4 ( \eta_1 -\eta_2)|_{\eta_1^3}\;
    \cO^c_{\Delta -\frac12}\Big(z_2, \bar z_2, \eta_2  \Big) ~,
    \ee
    where $|_{\eta_1}$ and $|_{\eta_1^3}$ means projection to terms with exactly one or three $\eta_1$'s.
So $\mathcal K ^a$ correponds to soft gluinos and we can rewrite 
\be
\mathcal K ^a(z,\bar z, \eta) =\eta^A   K_A ^a(z,\bar z)+\frac{1}{3!} \epsilon_{ABCD}\eta^A\eta^B\eta^C \bar {  K}^{D  \; a}(z,\bar z)~,
\ee
where 
\be
   K_A ^a(z,\bar z)=  \lim_{\Delta \to \frac12}(\Delta -\frac12) \sfGamma^a_{A\; \Delta}(z,\bar z)~, \qquad
 \bar {  K}^{A  \; a}(z,\bar z)=  \lim_{\Delta \to \frac12}(\Delta -\frac12) \bar\sfGamma ^{A\; a}_{ \Delta}(z,\bar z)~.
\ee
These soft gluinos are related to the soft gluons through supersymmetry. 

The conformal weights  $(h,\bar h)$  of $K$ and $\bar K$ are   $(1/2,0)$ and $(0,1/2)$, respectively. And we have
\beqn 
   K  ^a    (z_1, \bar z_1 )  O_{\Delta,J  }^b(z_2, \bar z_2 )
&\sim&
 \frac{f^{abc}}{z_{12}}     O^c_{ \Delta-\frac12,J-\frac12}\Big(z_2, \bar z_ 2   \Big)  ~,
\\
\bar{   K}  ^a    (z_1, \bar z_1 )  O_{\Delta,J  }^b(z_2, \bar z_2 )
&\sim&
 \frac{f^{abc}}{\bar z_{12}}   O^c_{ \Delta-\frac12,J+\frac12}\Big(z_2, \bar z_ 2   \Big)  ~,
  \eeqn
where we omit the $R$-symmetry indices for simplicity.

 \subsubsection{$\Delta\to 0$}
 
We then consider  the limit $\Delta  \to 0$ and define  
 \be
\mathcal L ^a(z, \bar z,\eta) =\lim_{\Delta \to 0} \Delta  \cO_{\Delta  }^a(z, \bar  z ,\eta )  ~.
\ee
 There are more complications in this case.  In \eqref{supeOPEsimple}, there are two sources of singularities in the limit $ {\Delta \to 0}$. In the first case, we allow exactly two  $e^{\frac12\p_{\Delta_1}}$ acting on    $B(\Delta_ 1-1,\Delta_2-1)$. This gives rise to the coefficient $B(\Delta_ 1 ,\Delta_2-1)=1/ \Delta_1 $ in the limit $\Delta_1\to 0$.
In the second case, there is no    $e^{\frac12\p_{\Delta_1}}$ acting on    $B(\Delta_ 1-1,\Delta_2-1)$:
\be
\lim_{\Delta_1 \to 0} B(\Delta_ 1-1,\Delta_2-1) =\frac{2-\Delta_2}{\Delta_1}~.
\ee

The   OPE \eqref{supeOPEsimple} then becomes
\beqn\label{OPELO}
\mathcal L ^a(z_1, \bar z_1,\eta_1) \cO_{\Delta_2   }^b(z_2, \bar z_2,\eta_2)
&\sim&
 \frac{f^{abc}}{z_{12}}     \cO^c_{\Delta_2  -1}\Big(z_2, \bar z_2, \eta_1  + \eta_2     \Big)  
\Big|_{\eta_1^2 }
+
 \frac{f^{abc}}{\bar z_{12}}       \delta^4 ( \eta_1 -\eta_2)|_{\eta_1^2}\;
    \cO^c_{\Delta_2 -1}\Big(z_2, \bar z_2, \eta_2  \Big) 
 \nonumber    \\&   &+
 \frac{f^{abc}}{z_{12}}     \cO^c_{\Delta_2 -1}\Big(z_2, \bar z_2, \eta_2   e^{ \frac12 \p_{\Delta_2}} \Big)  
 \Big(2- \Delta_2\Big)
 \nonumber
 \\&   &+
 \frac{f^{abc}}{\bar z_{12}}  
    \cO^c_{\Delta_2 -1}\Big(z_2, \bar z_2, \eta_2 e^{-  \frac12 \p_{\Delta_2}}  \Big) 
     \delta^4 ( \eta_1      )  
 \Big( - \Delta_2\Big)~.
    \eeqn
 In the above OPE, there are 2,2,0 and 4 $\eta_1$'s on the right hand side, respectively. So we can expand 
\be
\mathcal L ^a(z,\bar z, \eta) = \frac12 \eta^A \eta^B  L_{AB} ^a(z,\bar z)+
 L^a(z,\bar z)+ \bar L^a(z,\bar z)\delta^4(\eta )~,
\ee
where 
\be
  L_{AB} ^a(z,\bar z)=  \lim_{\Delta \to 0} \Delta  \sfPhi^a_{AB\; \Delta}(z,\bar z)~, \quad
   { L}^{  \; a}(z,\bar z)=  \lim_{\Delta \to 0}\Delta    \sfG ^{+ \; a}_{ \Delta}(z,\bar z)~, \quad
 L^{  \; a}(z,\bar z)=  \lim_{\Delta \to 0}\Delta    \sfG ^{- \; a}_{ \Delta}(z,\bar z)~.
\ee
They have weights $(h,\bar h)=(0,0),(1/2,-1/2),(-1/2,1/2)$, respectively.

Then the OPE \eqref{OPELO} decomposes into 
\beqn   \label{LABOope}
\frac12 \eta^A \eta^B  L_{AB} ^a(z_1, \bar z_1 ) \cO_{\Delta_2   }^b(z_2, \bar z_2,\eta_2)
\nonumber &\sim&
 \frac{f^{abc}}{z_{12}}     \cO^c_{\Delta_2  -1}\Big(z_2, \bar z_2, \eta_1  + \eta_2     \Big)  
\Big|_{\eta_1^2 }
\\&&+
 \frac{f^{abc}}{\bar z_{12}}       \delta^4 ( \eta_1 -\eta_2)|_{\eta_1^2}\;
    \cO^c_{\Delta_2 -1}\Big(z_2, \bar z_2, \eta_2  \Big) ~,
\\   \label{LbarOope}
L ^a(z_1, \bar z_1 ) \cO_{\Delta_2   }^b(z_2, \bar z_2,\eta_2)
&\sim&
 \frac{f^{abc}}{z_{12}}     \cO^c_{\Delta_2 -1}\Big(z_2, \bar z_2, \eta_2   e^{ \frac12 \p_{\Delta_2}} \Big)  
 \Big(2- \Delta_2\Big)~,
\\  \label{LOope}
\bar L ^a(z_1, \bar z_1 ) \cO_{\Delta_2   }^b(z_2, \bar z_2,\eta_2)
&\sim&
 \frac{f^{abc}}{\bar z_{12}}     \cO^c_{\Delta_2 -1}\Big(z_2, \bar z_2, \eta_2   e^{- \frac12 \p_{\Delta_2}} \Big)  
 \Big( - \Delta_2\Big)~.
\eeqn
 
To see closer, we also write down the component OPE after doing $\eta$ expansion. The OPE in \eqref{LABOope} becomes
 \be
L_{\bullet\bullet}^a(z,\bar z)   (z_1, \bar z_1 )  O_{\Delta,J  }^b(z_2, \bar z_2 )
 \sim 
 \frac{f^{abc}}{z_{12}}     \cO^c_{ \Delta-1,J-1}\Big(z_2, \bar z_ 2   \Big)  
 + \frac{f^{abc}}{\bar z_{12}}     \cO^c_{ \Delta-1,J+1}\Big(z_2, \bar z_ 2   \Big)  ~,
 \ee
 where we again ignore the $R$-symmetry index.  This corresponds to the soft scalar. 
 
 For \eqref{LOope} and \eqref{LbarOope}, the component OPEs  involving only gluons are:  
   \beqn
L ^a(z_1, \bar z_1 ) \sfG_{\Delta    }^{+b}(z_2, \bar z_2  )
&\sim&
 \frac{f^{abc}}{z_{12}}   (2- \Delta )    \sfG^{+c}_{\Delta -1}\Big(z_2, \bar z_2  \Big)  ~,
\\
L ^a(z_1, \bar z_1 ) \sfG_{\Delta   }^{-b}(z_2, \bar z_2  )
&\sim&
 \frac{f^{abc}}{z_{12}}   ( - \Delta )    \sfG^{-c}_{\Delta -1}\Big(z_2, \bar z_2  \Big)  ~,
\\
\bar L ^a(z_1, \bar z_1 ) \sfG_{\Delta   }^{+b}(z_2, \bar z_2  )
&\sim&
 \frac{f^{abc}}{\bar z_{12}}   ( - \Delta )    \sfG^{+c}_{\Delta -1}\Big(z_2, \bar z_2  \Big)  ~,
\\
\bar L ^a(z_1, \bar z_1 ) \sfG_{\Delta  }^{-b}(z_2, \bar z_2  )
&\sim&
 \frac{f^{abc}}{\bar z_{12}}   (2 - \Delta )    \sfG^{-c}_{\Delta -1}\Big(z_2, \bar z_2  \Big)  ~.
\eeqn
These OPEs   just  arise  from   the sub-leading soft gluon theorem.  Note that here we only derive the singular terms in the OPE. But for the sub-leading soft gluon theorem, there are actually also important phase factor terms $z_{12}/\bar z_{12}$ or $\bar z_{12}/  z_{12}$ which appears  in the Ward identity \eqref{KOWI}.   The OPE above agrees with the singular  term in \eqref{KOWI}. 
So it would be important to also find the non-singular sub-leading terms in the OPE by studying sub-leading collinear limit in SYM theory. We leave it to the future. 
 
 One can in principle continue the discussion to all $\Delta =-1/2, -1, -3/2, \cdots$. All correspond  to the conformally soft modes, giving rise to  various Ward identities.  For gluon and graviton, the symmetry algebra for the  infinite set of symmetries is recently determined in \cite{Guevara:2021abz}.
 
\section{Conclusion and outlook}\label{conclusion}

To summarize, in this paper, we have initiated  the study of  the celestial amplitude in $\mathcal N=4$ SYM theory using the language of  superfield.  We constructed the superconformal generators and the corresponding celestial on-shell superfield  living on the celestial sphere.  We then computed the three- and four-point celestial superamplitudes explicitly in  $\mathcal N=4$  SYM theory. 
We also studied the $\mathcal N=4$ SYM amplitude in the soft and collinear limit and their consequences in celestial conformal field theory. The energetically super-soft theorem in momentum space becomes the conformally super-soft theorem which further  leads to Ward identity of  symmetry current in CCFT. And the collinear limit enables us to extract the super-OPEs of  the  super-operators. 
 
In our discussion of     $\mathcal N=4$  SYM theory, we made full use of the superspace and on-shell superfield. The momentum space on-shell superspace techniques for theories with   $\mathcal N<4$  supersymmetry were studied in~\cite{Elvang:2011fx}. It is straightforward to generalize our celestial superspace to those theories with less supersymmetry. Therefore, our formalism provides a natural language to account for bulk supersymmetry and superconformal symmetry in CCFT. 

 In particular,  it would also be interesting to generalize our celestial  superfield techniques to   the supergravity theories, especially the $\mathcal N=8$  supergravity.~\footnote{Note that the  case with $\mathcal N=1$  supersymmetry   has already been considered in   component form without using superfield \cite{Fotopoulos:2020bqj}.} The asymptotic symmetries of those theories are given by the supersymmetric extension of BMS symmetry. The interplay between  supertranslation, superrotation, and   maximal  supersymmetry    is supposed to lead to a rich story on the celestial sphere. One could  furthermore consider the celestial double copy relation between  $\mathcal N=4$ SYM  and  $\mathcal N=8$  supergravity by generalizing the non-supersymmetric  celestial double copy in \cite{Casali:2020vuy}.

One reason for us to focus on $\mathcal N=4$ SYM theory in this paper is because of its maximal superconformal symmetry in 4D. However, a very remarkable feature of  $\mathcal N=4$ SYM theory  is that the     scattering amplitude  (in the planar limit) has been shown to enjoy much larger symmetries, the   Yangian symmetry, which can be regarded as the combination of superconformal and dual superconformal symmetry \cite{Drummond:2009fd}. 
As superconformal symmetry,   it is then natural to ask   how  to realize superconformal  and  Yangian symmetry   on the celestial sphere.  
Once all various types of symmetries are understood, a bootstrap program  for the CCFT of $\mathcal N=4$ SYM theory may become  feasible. 
A closely related question is how to  relate the celestial sphere with   twistor space where the $\mathcal N=4$ SYM theory was interpreted as B-model  topological string theory \cite{Witten:2003nn}.
 
 Another virtue of  $\mathcal N=4$ SYM theory is that the quantum corrections have been much more well-understood compared to pure YM theory.~\footnote{The infrared divergent loop amplitudes
in planar  $\mathcal N=4$  super Yang-Mills theory has been considered in \cite{Gonzalez:2020tpi}.}
  So it would be interesting to understand the fate of loop corrections in  celestial amplitude. This is important for further understanding   the full-fledged quantum gravity via  celestial holography.
 Even at the tree level, in the discussion of soft and collinear limits, we were mainly discussing the leading term. It would be  nice to work out the other sub-leading corrections.

Finally, a useful operation for celestial operator is shadow transformation \cite{Pasterski:2017kqt}. Therefore,   it would be interesting to see how to perform the shadow transformation for a super-operator. The shadow transformation flips the spin of the operator and in particular, should be compatible with supersymmetry. However, the supersymmetries considered here are 4D supersymmetry in the bulk instead of the ordinary 2D supersymmetry on the celestial sphere. So it seems to be not sensible  to combine $(z_1,\bar z_1, \eta_1^A)$ and $(z_2,\bar z_2, \eta_2^A)$ to form a distance that  could be used to construct the supershadow.  Nevertheless, it is worth studying whether one can  assemble    the shadow of each component operator  together  in a supersymmetric way.

 \acknowledgments

We would like to thank Rodolfo Russo and Congkao Wen for discussions  and comments on the manuscript.  
We also thank the authors of \cite{AndiCelestial}  for sharing their  draft before the publication, and Congkao Wen for informing us the two parallel independent works on   similar topics.
This work was  supported by the Royal Society grant, \textit{“Relations, Transformations, and Emergence in Quantum Field Theory”}, and 
by the Science and Technology Facilities Council (STFC) Consolidated Grant ST/T000686/1 \textit{``Amplitudes, strings  \& duality''}.


\appendix

\section{Soft and collinear limit in YM and SYM} \label{app}
\subsection{Soft theorem in YM and SYM}
 The soft theorem for (color-ordered) YM gluon amplitude is 
\be\label{YMsoftlimit}
\mathcal A_n ( \cdots, a, s^\pm , b, \cdots) \xrightarrow{p_s\to 0}  \Soft^{\text{YM}}  (a, s^\pm ,b) \mathcal A_n ( \cdots, a,  b, \cdots)~, 
\ee 
where  the soft factor for positive helicity soft gluon at tree level is given by \cite{He:2014bga}
%
\be\label{softfactor}
\Soft^{\text{YM}} (a, s^+ ,b) =\frac{\EV{ab}}{\EV{as}\EV{sb}}+
\Bigg( \frac{1}{\EV{sb} }\tilde \lambda_s^{\dot\alpha} \frac{\p}{\p \tilde\lambda_b^{\dot\alpha}}
+ \frac{1}{\EV{as} }\tilde \lambda_s^{\dot\alpha} \frac{\p}{\p \tilde\lambda_a^{\dot\alpha}}
\Bigg)~.
\ee
The first term is the leading soft factor  of order $\mathcal O(1/p_s)$, while the second term in parentheses   is  the  subleading soft factor of order $\mathcal O(1/p_s^0)$. For negative helicity soft gluon, the soft factor is given by the conjugate $\lambda_i \leftrightarrow \tilde \lambda_i$.

The discussion can be generalized to the superamplitude in $\mathcal N=4$ SYM theory:
\be\label{SYMsofttheorem}
\mathscr A_n ( \cdots, a, s  , b, \cdots) \xrightarrow{p_s\to 0}  \Soft^{\text{SYM}} (a, s  ,b) \mathscr  A_{n-1} ( \cdots, a,  b, \cdots) ~,
\ee 
where each particle is labeled by $(\lambda_i, \tilde\lambda_i, \eta_i)$. 
 The physical soft limit $p_s\to 0$    corresponds to setting  $(\lambda_s, \tilde \lambda_s, \eta_s) \to (\epsilon\Lambda_s, \epsilon\tilde \Lambda_s, \eta_s)  $ and then taking $\epsilon \to 0$.  However, this is a little complicated.   In the literature,  one instead  considers  the holomorphic soft limit $(\lambda_s, \tilde \lambda_s, \eta_s) \to (\epsilon\Lambda_s, \tilde \lambda_s, \eta_s)  $ and anti-holomorphic  soft limit $(\lambda_s, \tilde \lambda_s, \eta_s) \to ( \lambda_s, \epsilon\tilde \Lambda_s, \eta_s) $.

In the holomorphic soft limit $(\lambda_s, \tilde \lambda_s, \eta_s) \to (\epsilon\Lambda_s, \tilde \lambda_s, \eta_s)  $, namely $\lambda_s\to0 $, the supersoft factor is given by \cite{He:2014bga}
\be
\Soft^{\text{SYM}}_{\text{hol}} (a, s  ,b) =\frac{\EV{ab}}{\EV{as}\EV{sb}}+
\Bigg[ \frac{1}{\EV{sb} } \Big(\tilde \lambda_s^{\dot\alpha}    \frac{\p}{\p \tilde\lambda_b^{\dot\alpha}} 
+ \eta_s^A \frac{\p}{\p \eta_b^A}\Big)
+ \frac{1}{\EV{as}  } \Big( \tilde \lambda_s^{\dot\alpha} \frac{\p}{\p \tilde\lambda_a^{\dot\alpha}}
+  \eta_s^A \frac{\p}{\p \eta_a^A}\Big) 
\Bigg]~,
\ee
while  in the anti-holomorphic soft limit $(\lambda_s, \tilde \lambda_s, \eta_s) \to ( \lambda_s,\epsilon \tilde \Lambda_s, \eta_s)  $, namely $\tilde \lambda_s\to0 $, the supersoft factor is given by \cite{He:2014bga}
\be
\Soft^{\text{SYM}}_{\text{anti-hol}}  (a, s  ,b) =\frac{\SV{ab}}{\EV{as}\SV{sb}}\delta^4 \Big( \eta_s +  \frac{[as]}{[ab]} \eta_b+  \frac{[ sb]}{[ab]} \eta_a\Big)
\Bigg[ 1+ \Big(
 \frac{\SV{sb}}{\SV{ab} }  \eta_s^A \frac{\p}{\p \eta_b^A} +
 \frac{\SV{as}}{\SV{ab} }  \eta_s^A \frac{\p}{\p \eta_a^A} 
 \Big)
\Bigg]~.
\ee
In these two limits above, we give  both the leading and sub-leading soft factors.

In the physical soft limit $p_s\to 0$ or equivalently $\lambda_s,\tilde\lambda_s\to0 $, the leading     soft factor is given by the sum of leading soft factors in holomorphic and anti-holomorphic soft limit: 
\be\label{supersoftfactor}
\Soft_{\text{leading}}^{\text{SYM}} (a, s  ,b) =\frac{\EV{ab}}{\EV{as}\EV{sb}}+
\frac{\SV{ab}}{\SV{as}\SV{sb}}\delta^4 (\eta_s )~.
\ee
The two terms obviously correspond to the soft gluons of positive and negative helicity. Other components are absent here because their corresponding soft factors are sub-dominant (for example, the leading soft factor for gluinos scales as $1/\sqrt{p_s}$ \cite{Fotopoulos:2020bqj}). 

\subsection{Collinear limit in  YM}
When two particles point to the same direction, the amplitude also becomes   singular, knowns as the collinear singularity. More specifically, consider two particles with momenta $p_1$ and $p_2$. The two particles can fuse into one particle with momentum $P_{12}\equiv p_1+p_2$ and one can parametrize the collinear momenta as $p_1=\sfz P_{12} , p_2=(1-\sfz)P_{12}$. In the  collinear limit, the gluon amplitude  in YM theory satisfies 
\be\label{YMcollinear}
\mathcal A_n (1^{h_1}, 2^{h_2}, \cdots ) \xrightarrow{p_1\slash\!\slash p_2} 
\sum_h\mathcal  A_{n-1} (P_{12}^h\; , \cdots ) \Split_{-h} (1^{h_1}, 2^{h_2})~,
\ee
where the split factors  $ \Split_{-h} (1^{h_1}, 2^{h_2}) $ are given by  \cite{Henn:2014yza,Ferro:2020lgp}
%
\beqn
\Split_+ (\sfz; 1^+, 2^-) &=&\frac{(1-\sfz)^2}{\sqrt{\sfz (1-\sfz)}}\frac{1}{\EV{12}}~,
   \\
\Split_+ (\sfz; 1^-, 2^+) &=&\frac{ \sfz^2}{\sqrt{\sfz (1-\sfz)}}\frac{1}{\EV{12}} ~,
   \\ \label{splitYM}
\Split_+ (\sfz; 1^-, 2^-) &=&\frac{1}{\sqrt{\sfz (1-\sfz)}}\frac{1}{\SV{12}} ~,
    \\
\Split_+ (\sfz; 1^+, 2^+) &=&0    ~,
\eeqn
and
\be\label{YMsplitfactor}
\Split_- (\sfz; 1^{-h_1}, 2^{-h_2})=
\Split_+ (\sfz; 1^{h_1}, 2^{h_2})|_{\EV{12}\leftrightarrow \SV{12}}~.
\ee

 One can actually derive the soft theorem by  taking the consecutive collinear limits of the same type  \cite{Ferro:2020lgp}. 
 The soft factor derived in this way  is in agreement with the leading   term in \eqref{softfactor} as well as its conjugate.

\subsection{Collinear limit in SYM}
We can also consider the collinear limit of SYM theory using the language of superamplitude. Here we follow the discussion in \cite{Ferro:2020lgp}. 
The SYM amplitude  can be written as
\be
\mathscr  A_n=A_{n,2}+A_{n,4}+\cdots +A_{n,n-2}~,
\ee
which corresponds to MHV, NMHV,..., $\overline  {\text{MHV}}$. 
In the collinear limit, the SYM amplitude satisfies
\be\label{supersplitprocess}
A_{n,k}(1,2,3,\cdots, n) \xrightarrow{P_{12}^2\to 0} \sum_{l=1}^2 \int d^4  \eta_{P_{12}} \; \Split_{1-l}(1,2,P_{12})  A_{n-1,k-l+1}(P_{12},3, \cdots ,n)~,
\ee
where $l=1,2$ corresponds to helicity preserving and helicity decreasing process.  Summing over $k$ on both sides  gives
\be\label{SYMcollinear}
\mathscr A_n(1,2,3,\cdots, n) \xrightarrow{P_{12}^2\to 0} \sum_{l=1}^2 \int d^4  \eta_{P_{12}}\; \Split_{1-l}(1,2,P_{12})  \mathscr A_{n-1}(P_{12},3, \cdots ,n)~.
\ee

For helicity preserving case, the super-split factor is given by \cite{Ferro:2020lgp}
\be\label{SYMsplit1}
\Split_0 (\sfz; \eta_1,\eta_2, \eta_3)=\frac{1}{\sqrt{\sfz(1-\sfz)}}\frac{1}{\EV{12}}\prod_{A=1}^4 \Big(\eta_3^A -\sqrt\sfz \eta_1^A -\sqrt{1-\sfz} \eta_2^A \Big)~.
\ee
While for helicity decreasing case, the super-split factor is given by \cite{Ferro:2020lgp}
\be\label{SYMsplit2}
\Split_{-1} (\sfz; \eta_1,\eta_2, \eta_3)=\frac{1}{\sqrt{\sfz(1-\sfz)}}\frac{1}{\SV{12}}\prod_{A=1}^4 \Big(\eta_1^A \eta_2^A - \sqrt{1-\sfz} \eta_1^A \eta_3^A  +  \sqrt{ \sfz} \eta_2^A \eta_3^A\Big)~.
\ee
 
As in the pure Yang-Mills theory, one  can also  produce the  soft limit by taking consecutive collinear limits. Taking $p_2\to 0$ and thus $\sfz\to 1$, the split factors reduce to 
\beqn
\Split_0 (\sfz; \eta_1,\eta_2, \eta_{P_{12}}) &\to&
 \frac{\EV{13}} {\EV{12}\EV{23}}  \prod_{A=1}^4  ( \eta_{P_{12}}^A- \eta_1 ^A) ~,
\\
\Split_{-1} (\sfz; \eta_1,\eta_2, \eta_{P_{12}})
&\to&
 \frac{\SV{13}} {\SV{12}\SV{23}}
\prod_{A=1}^4  \eta_2^A( \eta_{P_{12}}^A- \eta_1 ^A) ~.
\eeqn
Inserting into \eqref{supersplitprocess}, one obtains 
\beqn
A_{n,k}(1,2,3,\cdots, n) &\xrightarrow {p_2\to 0} & \frac{\EV{13}} {\EV{12}\EV{23}}A_{n-1,k}(1, 3,\cdots, n)~,
 \\
A_{n,k}(1,2,3,\cdots, n) &\xrightarrow {p_2\to 0} &\delta^4(\eta_2) \frac{\SV{13}} {\SV{12}\SV{23}}A_{n-1,k-1}(1, 3,\cdots, n)~.
\eeqn

Combining together, we have
\be
\mathscr A_{n }(1,2,3,\cdots, n)  \xrightarrow {p_2\to 0} \Big(  \frac{\EV{13}} {\EV{12}\EV{23}}+  
\frac{\SV{13}} {\SV{12}\SV{23}} \delta^4(\eta_2)\Big)   \mathscr A_{n-1 }(1, 3,\cdots, n) ~.
\ee
This is in agreement with the leading supersoft factor in \eqref{supersoftfactor}.


\appendix

 \bibliographystyle{JHEP} 
 
 \bibliography{BMS}
 
\end{document}